\documentclass[prd,onecolumn,reprint,superscriptaddress]{revtex4-2}
\usepackage{feynmp,amsmath,mathtools,amssymb,graphicx,subfigure,hyperref,bm,mathrsfs,slashed}
\usepackage{CJK}
\usepackage{color,ulem}

\begin{document}
		\begin{CJK}{UTF8}{song}
		
		\title{Highly anisotropic optical conductivities in two-dimensional tilted semi-Dirac bands}
		
		\author{Chang-Xu Yan}
		\thanks{These authors contributed equally to this work.}
		\affiliation{Department of Physics, Institute of Solid State Physics and
			Center for Computational Sciences, Sichuan Normal University, Chengdu,
			Sichuan 610066, China}
        \affiliation{Department of Physics, Beijing Normal University, Beijing 100875, China}
		
		\author{Chao-Yang Tan}
		\thanks{These authors contributed equally to this work.}
		\affiliation{Department of Physics, Institute of Solid State Physics and
			Center for Computational Sciences, Sichuan Normal University, Chengdu,
			Sichuan 610066, China}
        \affiliation{Department of Physics and Beijing Key Laboratory of Opto-electronic Functional Materials and Micro-nano Devices, Renmin University of China, Beijing 100872, China}
		
		\author{Hong Guo}
		\affiliation{Department of Physics, McGill University, Montreal, Quebec H3A 2T8, Canada}
		\affiliation{Department of Physics, Institute of Solid State Physics and Center for Computational Sciences,
			Sichuan Normal University, Chengdu, Sichuan 610066, China}

		\author{Hao-Ran Chang}
		\thanks{Corresponding author:hrchang@mail.ustc.edu.cn}
		\affiliation{Department of Physics, Institute of Solid State Physics and Center for Computational Sciences, Sichuan Normal University, Chengdu, Sichuan 610066, China}
		
		\date{\today}
		
		\begin{abstract}
			Within linear response theory, the absorptive part of highly anisotropic optical conductivities are analytically calculated for distinct tilts in two-dimensional (2D) tilted semi-Dirac bands (SDBs). The transverse optical conductivities always vanish.
			The interband longitudinal optical conductivities (LOCs) in 2D tilted SDBs differ qualitatively in the power-law scaling of $\omega$
			as $\mathrm{Re}\sigma_{\perp}^{\mathrm{IB}}(\omega)\propto\sigma_0\sqrt{\omega}$ and $\mathrm{Re}\sigma_{\parallel}^{\mathrm{IB}}(\omega)\propto\sigma_0
			/\sqrt{\omega}$. By contrast, the intraband LOCs in 2D tilted SDBs depend on $\mu$ in the power-law scaling as
			$\mathrm{Re}\sigma_{\perp}^{\mathrm{D}}(\omega)\propto\sigma_0\mu \sqrt{\mu}$ and $\mathrm{Re}\sigma_{\parallel}^{\mathrm{D}}(\omega)\propto\sigma_0
			\mu/\sqrt{\mu}$. The tilt-dependent behaviors of LOCs could qualitatively characterize distinct impact of band tilting in 2D tilted SDBs. In particular, for arbitrary tilt $t$ satisfying $0<t\le 2$, the interband LOCs always possess a robust fixed point at $\omega=2\mu$. The power-law scalings and tilt-dependent behaviors further dictate significant differences
in the asymptotic background values and angular dependence of LOCs. Our theoretical predictions
should be valid for a broad class of 2D tilted SDB materials, and can also be used to fingerprint 2D
tilted SDB from 2D untilted SDB as well as tilted Dirac bands.
		\end{abstract}

		\maketitle
	\end{CJK}

	\section{Introduction\label{Sec:intro}}
	
	The anisotropy of energy dispersion usually dictate interesting behaviors of anisotropic physical quantities. In ordinary semiconductors, the anisotropic dispersions are mainly reflected by respective effective masses along different spatial directions, while in gapless Dirac materials, the effective mass is always zero. Interestingly, a kind of exotic energy dispersion has been predicted in a series of two-dimensional (2D) materials, such as $\rm{VO_2/TO_2}$ heterostructures \cite{PRLPardo2009,PRLBanerjee2009}, (BEDT-TTF)$_2$I$_3$ salt under pressure \cite{JPSJ2006}, photonic crystals \cite{OEPYWu2014} and strained honeycomb lattices \cite{NJPWunsch2008,PRBMontambaux2009,PRBHasegawa2006,PRLZhu2007}, whose low energy dispersion around Dirac points is linear in one spatial direction but quadratic in the perpendicular direction. This 2D hybrid dispersion is termed semi-Dirac bands (SDBs) possessing an intrinsic anisotropy, and has also been experimentally realized in black phosphorous \cite{SCIKim2015}. 
	
	Due to the hybrid dispersion, the semi-Dirac materials anticipate anisotropic transport properties, such as hydrodynamic transport properties \cite{PRLKim2018},  magnetoconductivity \cite{PBLZhou2021}, and optical conductivity \cite{PRXRoy2018,PRBCarbotte2019,PRBCarbotte2019B,PRBZhang2022,PRBOriekhov2022}. It has been further predicted that 2D SDBs can be tilted along a specific direction in momentum space \cite{PCLZhang2017}. As a result, different phases of Lifshitz transition can exhibit in 2D tilted SDBs, just as that in 2D tilted Dirac bands \cite{Lifshitz1960,Volovik2017,Volovik2018}. Interestingly, band tilting in 2D tilted Dirac bands can remarkably enhance the anisotropy of physical quantities, and the resulting Lifshitz transition leads to significantly different properties including plasmons \cite{JPSNishine2011,JPSNishine2010,PRBIurov2017,PRBAgarwal2017,PRBJafari2018,PRBMojarro2022,arXivYan2022} and optical conductivities \cite{JPSNishine2010,PRBMojarro2022,PRBVerma2017,PRBHerrera2019,PRBRostamzadeh2019,PRBTan2021,PRBTan2022,arXivHou2022,PRBMojarro2021,PRBYao2021,PRBIurov2018,PRBIurov2020}. It can therefore be expected that the semi-Dirac bands together with band tilting can lead to highly anisotropic and tilting-dependent behaviors of transport properties. However, the relevant transport studies have been restricted to 2D untilted SDBs so far \cite{PRLKim2018,PBLZhou2021,PRXRoy2018,PRBCarbotte2019,PRBZhang2022,PRBOriekhov2022}, and no transport study is available for 2D tilted SDBs.
	
	As a typical transport quantity, optical conductivity can be used to extract the essential information of band structure. Specifically, the exotic behaviors of 
	optical conductivity can be used to characterize
	the Lifshitz transition in 2D tilted Dirac bands \cite{PRBTan2021}. To show its characteristic behaviors and the qualitative differences and similarities compared with that in 2D untilted SDBs and 2D tilted Dirac bands, we theoretically study the optical conductivities in 2D tilted SDBs. The rest of the paper is organized as follows. In Sec. \ref{Sec:Theoretical formalism}, we introduce the Hamiltonian and theoretical formalism. In Sec. \ref{Sec:Interband conductivity}, we reveal the effect of band tilting by elaborating results for interband optical conductivity. Meanwhile, we  report that the power-law scaling of optical conductivity in 2D untilted semi-Dirac material is universal for different band tilting. In addition, the intraband conductivities are analytically calculated in Sec. \ref{Sec:Intraband conductivity}. The discussion and summary are given in Sec. \ref{Sec: Discussion and Summary}. Finally, we present appendices to show the detailed calculation and analysis.

	\section{Model and Theoretical formalism \label{Sec:Theoretical formalism}}
	
	We begin with the effective Hamiltonian in the vicinity of a pair of Dirac points for 2D tilted SDBs
	\begin{align}
		\mathcal{H}_\kappa(k_x,k_y)=\kappa \hbar v_t k_y\tau_0+a \hbar^2 k_x^2\tau_1+\hbar v_F k_y\tau_2,
		\label{Eq1}
	\end{align}
	where $\kappa=\pm$ labels two valleys around Dirac points, $\boldsymbol{k}=(k_x,k_y)$ stands for the
	wave vector, $\tau_0$ and $\tau_i$ denote the $2\times2$ unit matrix and Pauli matrices, respectively. Distinct
	from both linearities along $k_x$ and $k_y$ in 2D Dirac bands, the dispersion here is linear in $k_y$ but quadratic
	in $k_x$. The parameter $a$ represents the inverse of $2m^{\ast}$ with $m^{\ast}$
	the effective mass, and $v_F$ and $v_t$ denote the Fermi velocity along $k_y$ and band tilting, respectively. For simplicity, we hereafter set $\hbar=1$ and define the tilt parameter
	\begin{align}
		t=\frac{v_t}{v_F}
	\end{align}
	to characterize band tilting for convenience.
	
	The eigenvalue of the Hamiltonian reads
	\begin{align}
		\varepsilon_\kappa^\lambda(\boldsymbol{k})=\varepsilon_\kappa^\lambda(k_x,k_y)=\kappa t v_F k_y+\lambda \mathcal{Z}(k_x,k_y),
	\end{align}
	where
	\begin{align}
		\mathcal{Z}(k_x,k_y)=\sqrt{\left(ak_x^2\right)^2+v_F^2 k_y^2},
	\end{align}
	and $\lambda=+$ and $\lambda=-$ denote the conduction band and valence band, respectively. It is noted that the 2D tilted SDBs $\varepsilon_{\kappa}^\lambda(k_x,k_y)$ remains invariant under the transformation $(-k_x,-k_y,+\kappa)\leftrightarrow(+k_x,+k_y,-\kappa)$ due to time reversal symmetry. The 2D tilted SDBs can also be categorized into four distinct ``phases" via the tilt parameter $t$, namely the untilted phase ($t=0$), type-I phase ($0<t<1$), type-II phase ($t>1$), and
	type-III phase ($t=1$), similar as that in the tilted Dirac bands \cite{Volovik2017,Volovik2018,PRBWild2022,PRBTan2022}.

	\begin{figure*}[ht]
		\includegraphics[width=14cm]{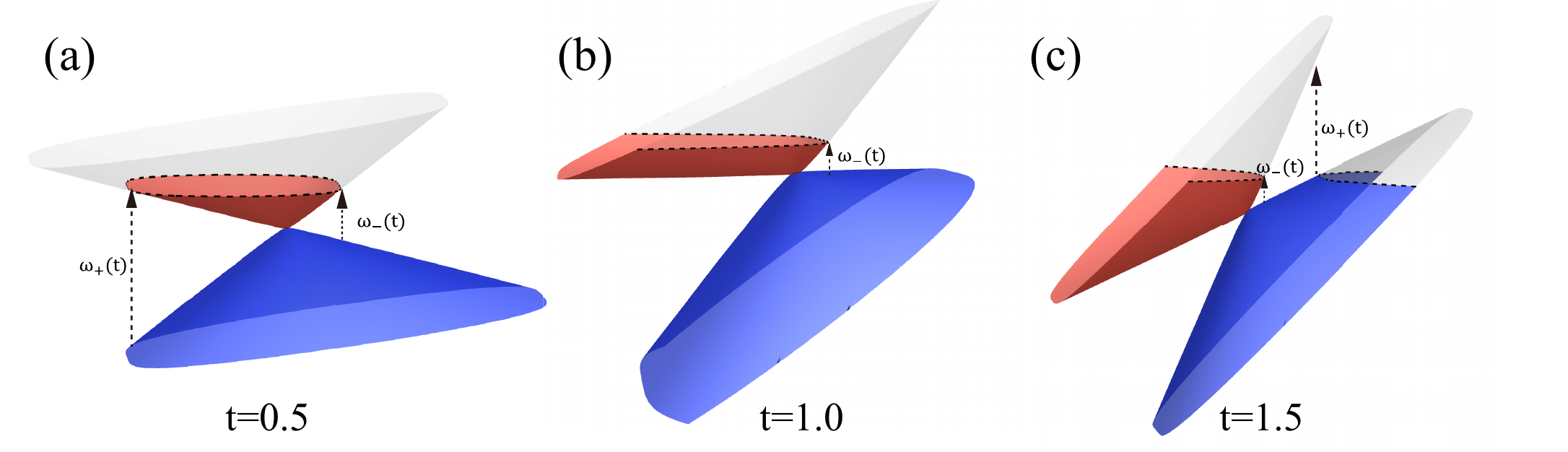}
		\caption{Energy bands and the corresponding critical frequencies $\omega_{\pm}(t)$ of interband optical transitions for distinct tilt $t$.}
		\label{distinct_tilt}
	\end{figure*}	
	
	Within the linear response theory, the optical conductivity is given by
	\begin{align}
		\sigma_{ij}(\omega,\mu,t)&=\sum_{\kappa=\pm}\sigma_{ij}^\kappa(\omega,\mu,t),
	\end{align}
	where $\omega$ denotes the photon frequency, $\mu$ measures the chemical potential with respect to the Dirac point, and $i,j$ represent the spatial components $x,y$. It is easy to verify that $\sigma_{ij}(\omega,\mu,t)$ respects the particle-hole symmetry, namely,
	\begin{align}
		\sigma_{ij}(\omega,\mu,t)&=\sigma_{ij}(\omega,-\mu,t)=\sigma_{ij}(\omega,|\mu|,t).
	\end{align}

	\begin{widetext}
	In general, the optical conductivity at the valley $\kappa$ is written as \cite{arXivHou2022} 
		\begin{align}
			\sigma_{ij}^{\kappa}(\omega,\mu,t)=\frac{\text{i}}{\omega}\lim_{\boldsymbol{q}\to\boldsymbol{0}}
			\int_{-\infty}^{+\infty}\frac{\mathrm{d}^{2}\boldsymbol{k}}{(2\pi)^{2}}\sum_{\lambda,\lambda^{\prime}=\pm}
			\mathcal{F}_{ij}^{\kappa;\lambda\lambda^{\prime}}[\boldsymbol{k};\boldsymbol{k}+\boldsymbol{q}]
			\frac{f[\varepsilon^{\lambda}_{\kappa}(\boldsymbol{k})]-f[\varepsilon_{\kappa}^{\lambda^{\prime}}(\boldsymbol{k}+\boldsymbol{q})]}{
\omega+\varepsilon^{\lambda}_{\kappa}(\boldsymbol{k})
-\varepsilon_{\kappa}^{\lambda^{\prime}}(\boldsymbol{k}+\boldsymbol{q})+\text{i}\eta}.
		\end{align}
		In the following, we focus on the real part of $\sigma^{\kappa}_{ij}(\omega,\mu,t)$  which can be divided into interband (IB) part and intraband (D) part as
		\begin{align}
			\mathrm{Re}\sigma_{ij}^{\kappa}(\omega,\mu,t)
			&=\mathrm{Re}\sigma_{ij(\mathrm{IB})}^\kappa(\omega,\mu,t)+\mathrm{Re}\sigma_{ij(\mathrm{D})}^\kappa(\omega,\mu,t)\notag\\
			&=\begin{cases}
				\mathrm{Re}\sigma_{ij(\mathrm{IB})}^{\kappa}(\omega,\mu,t)+\Theta[\mu]\mathrm{Re}\sigma_{ij(\mathrm{D})}^{\kappa,+}(\omega,\mu,t)
				+\Theta[-\mu]\mathrm{Re}\sigma_{ij(\mathrm{D})}^{\kappa,-}(\omega,\mu,t), & 0\leq t \leq1,\\\\
				\mathrm{Re}\sigma_{ij(\mathrm{IB})}^{\kappa}(\omega,\mu,t)+\mathrm{Re}\sigma_{ij(\mathrm{D})}^{\kappa,+}(\omega,\mu,t)
				+\mathrm{Re}\sigma_{ij(\mathrm{D})}^{\kappa,-}(\omega,\mu,t), & t>1,
			\end{cases}
			\label{LOCs}
		\end{align}
		where $\Theta(x)$ is the Heaviside step function satisfying $\Theta(x)=0$ for $x\le0$ and $\Theta(x)=1$ for $x>0$. The interband and intraband conductivities are given respectively as
		\begin{align}
			\mathrm{Re}\sigma_{ij(\mathrm{IB})}^\kappa(\omega,\mu,t)
			&=\pi \int^{+\infty}_{-\infty}\frac{dk_x}{2\pi} \int^{+\infty}_{-\infty}\frac{dk_y}{2\pi}\mathcal{F}^{\kappa;-+}_{ij}(k_x,k_y)\frac{ f\left[\varepsilon_\kappa^{-}(k_x,k_y)\right] -f\left[\varepsilon_\kappa^{+}(k_x,k_y)\right]}{\omega}\delta\left[\omega-2\mathcal{Z}(k_x,k_y)\right],\label{interLOC}\\
			\mathrm{Re}\sigma_{ij(\mathrm{D})}^{\kappa,\lambda}(\omega,\mu,t)
			&=\pi\int^{+\infty}_{-\infty}\frac{dk_x}{2\pi} \int^{+\infty}_{-\infty}\frac{dk_y}{2\pi} \mathcal{F}^{\kappa;\lambda\lambda}_{ij}(k_x,k_y) \left[-\frac{d f\left[\varepsilon_\kappa^{\lambda}(k_x,k_y),\mu\right]} {d\varepsilon_\kappa^{\lambda}(k_x,k_y)}\right]\delta(\omega),
			\label{intraLOC}
		\end{align}
		where $f(x)=\left\{1+\exp[(x-\mu)/k_BT]\right\}^{-1}$ is the Fermi distribution function, and the explicit expressions of $\mathcal{F}_{ij}^{\kappa;\lambda\lambda^{\prime}}(k_x,k_y)$ are given by
\begin{align}
&\mathcal{F}_{ij}^{\kappa;\lambda\lambda^{\prime}}(k_x,k_y)=
\lim_{\boldsymbol{q}\to \boldsymbol{0}}
\mathcal{F}_{ij}^{\kappa;\lambda\lambda^{\prime}}[\boldsymbol{k};\boldsymbol{k}+\boldsymbol{q}]
\nonumber\\&=
\begin{cases}
\mathcal{F}_{xy}^{\kappa;\lambda\lambda^{\prime}}(k_x,k_y)
=\mathcal{F}_{yx}^{\kappa;\lambda\lambda^{\prime}}(k_x,k_y)
=e^2\left[(\lambda+\lambda^{\prime})\frac{\kappa a^2 v_{t}k_x^3}{\mathcal{Z}(k_x,k_y)}
+2\lambda\lambda^{\prime}\frac{a^2 v_F^2k_x^3k_y}{[\mathcal{Z}(k_x,k_y)]^2}\right],
& (i,j)=(x,y)~\mathrm{or}~(y,x),\\
\mathcal{F}_{xx}^{\kappa;\lambda\lambda^{\prime}}(k_x,k_y) =2e^2a^2k_x^2\left[1+\lambda\lambda^{\prime}\frac{\left(ak_x^2\right)^2-v_F^2k_y^2}{\left[\mathcal{Z}(k_x,k_y)\right]^2}\right],
& (i,j)=(x,x),\\
\mathcal{F}_{yy}^{\kappa;\lambda\lambda^{\prime}}(k_x,k_y)=\frac{e^2}{2}\left\{\left[2v_t^2+4\kappa\lambda v_F^2\frac{v_tk_y}{\mathcal{Z}(k_x,k_y)}\right]\delta_{\lambda\lambda^{\prime}} +v_F^2\left[1-\lambda\lambda^{\prime}\frac{\left(ak_x^2\right)^2-v_F^2k_y^2}{\left[\mathcal{Z}(k_x,k_y)\right]^2}\right]\right\},
& (i,j)=(y,y),\\
\end{cases}
\label{Fexpr}
\end{align}
with $\delta_{\lambda\lambda^\prime}$ the Kronecker symbol. More details can be found in Appendix \ref{App1}.
\end{widetext}

Basically, the Hall conductivity and transverse optical conductivities are related to the Berry curvature and the resulting Chern number. For the 2D untilted ($t=0$) SDBs with topologically non-trivial Chern number ($C\neq0$), the Hall conductivity and transverse optical conductivities do not vanish \cite{PRBHuaqing2015,PRBQing-Yun2023}. However, for the 2D tilted ($t\neq 0$) SDBs in this work, the Chern number is always topologically trivial ($C=0$), because the Berry curvature is an odd function of $k_x$ for arbitrary band tilting parameter. This trivial topology leads to the vanished transverse optical conductivities, namely, $\mathrm{Re}\sigma_{xy}(\omega,\mu,t)=\mathrm{Re}\sigma_{yx}(\omega,\mu,t)=0$, similar to that in the 2D untilted ($t=0$) SDBs with topologically trivial Chern number ($C=0$) \cite{PRBCarbotte2019}. This result can be confirmed by an explicit analysis of the interband and intraband conductivities in Eqs. (\ref{interLOC}) and (\ref{intraLOC}), based on two relations that $\mathcal{F}_{xy}^{\kappa;\lambda\lambda^{\prime}}(-k_x,k_y)=-\mathcal{F}_{xy}^{\kappa;\lambda\lambda^{\prime}}(+k_x,k_y)$ and $\varepsilon_\kappa^\lambda(-k_x,k_y)=\varepsilon_\kappa^\lambda(+k_x,k_y)$ in the symmetric integration region of $k_{x}$.

Hereafter, we turn to the real part of longitudinal optical conductivities (LOCs) and restrict our analysis to the case with $\mu\ge0$. After summing over the contribution of two valleys, the LOC in Eq.(\ref{LOCs}) can be recast as
		\begin{align} \mathrm{Re}\sigma_{jj}(\omega,\mu,t)&=\mathrm{Re}\sigma_{jj}^{\mathrm{IB}}(\omega,\mu,t)+\mathrm{Re}\sigma_{jj}^{\mathrm{D}}(\omega,\mu,t).
			\label{LOCsp}
		\end{align}
To analytically calculate the interband LOCs $\mathrm{Re}\sigma_{jj}^{\mathrm{IB}}(\omega,\mu,t)$ and intraband LOCs $\mathrm{Re}\sigma_{jj}^{\mathrm{D}}(\omega,\mu,t)$ for different band tilting, we assume zero temperature $T\to0~\mathrm{K}$ where the Fermi distribution function $f(x)$ reduces to Heaviside step function $\Theta(\mu-x)$.

	\section{INTERBAND CONDUCTIVTY
		\label{Sec:Interband conductivity}}
	The interband LOCs $\mathrm{Re}\sigma_{jj}^{\mathrm{IB}}(\omega,\mu,t)$ can be further decomposed into two parts
\begin{align}
&\mathrm{Re}\sigma_{jj}^{\mathrm{IB}}(\omega,\mu,t)
\nonumber\\&
=\mathrm{Re}\sigma_{\perp}^{\mathrm{IB}}(\omega,\mu,t)\delta_{jx}
		+\mathrm{Re}\sigma_{\parallel}^{\mathrm{IB}}(\omega,\mu,t)\delta_{jy},
	\end{align}
	where
	\begin{align} \mathrm{Re}\sigma_{\chi}^{\mathrm{IB}}(\omega,\mu,t)=&N_{f}S_{\chi}^{(\mathrm{IB})}(\omega)\Gamma_{\chi}^{(\mathrm{IB})}(\omega,\mu,t),
		\label{LOCsp1}
	\end{align}
	with the index $\chi=\perp,\parallel$ denoting the spatial direction with respect to tilting direction and the degeneracy factor $N_{f}=g_{s}g_{v}$ accounting for spin degeneracy $g_s=2$ and valley degeneracy $g_v=2$. More calculation details of $S_{\chi}^{(\mathrm{IB})}(\omega)$ and $\Gamma_{\chi}^{(\mathrm{IB})}(\omega,\mu,t)$ are found in Appendix \ref{App2}. Two remarks on the interband LOCs are in order here. First, the dimensional auxiliary function $S_{\chi}^{(\mathrm{IB})}(\omega)$ in Eq.(\ref{LOCsp1}) can be explicitly written as
	\begin{align}
		S_{\perp}^{(\mathrm{IB})}(\omega)&=\frac{\sigma_{0}\sqrt{2a \omega}}{2\pi\upsilon_F},\\
		S_{\parallel}^{(\mathrm{IB})}(\omega)&=\frac{\sigma_{0}v_F}{2\pi\sqrt{2a\omega}},
	\end{align}
	where $\sigma_{0}=\frac{e^2}{4\hbar}$ (we restore $\hbar$ temporarily for explicitness). Evidently, the product $S_{\perp}^{(\mathrm{IB})}(\omega)\times S_{\parallel}^{(\mathrm{IB})}(\omega)=\frac{\sigma_{0}^2}{4\pi^2}$ is a constant independent of $\omega$. Second, the auxiliary function $\Gamma_{\chi}^{(\mathrm{IB})}(\omega,\mu,t)$ depends on $\omega$ and $\mu$ in terms of their ratio $\omega/\mu$, and hence is dimensionless. The explicit expressions of $\Gamma_{\chi}^{(\mathrm{IB})}(\omega,\mu,t)$ for distinct $t$ are listed in the following two subsections.

	\begin{figure*}[tbp]
		\includegraphics[width=10cm]{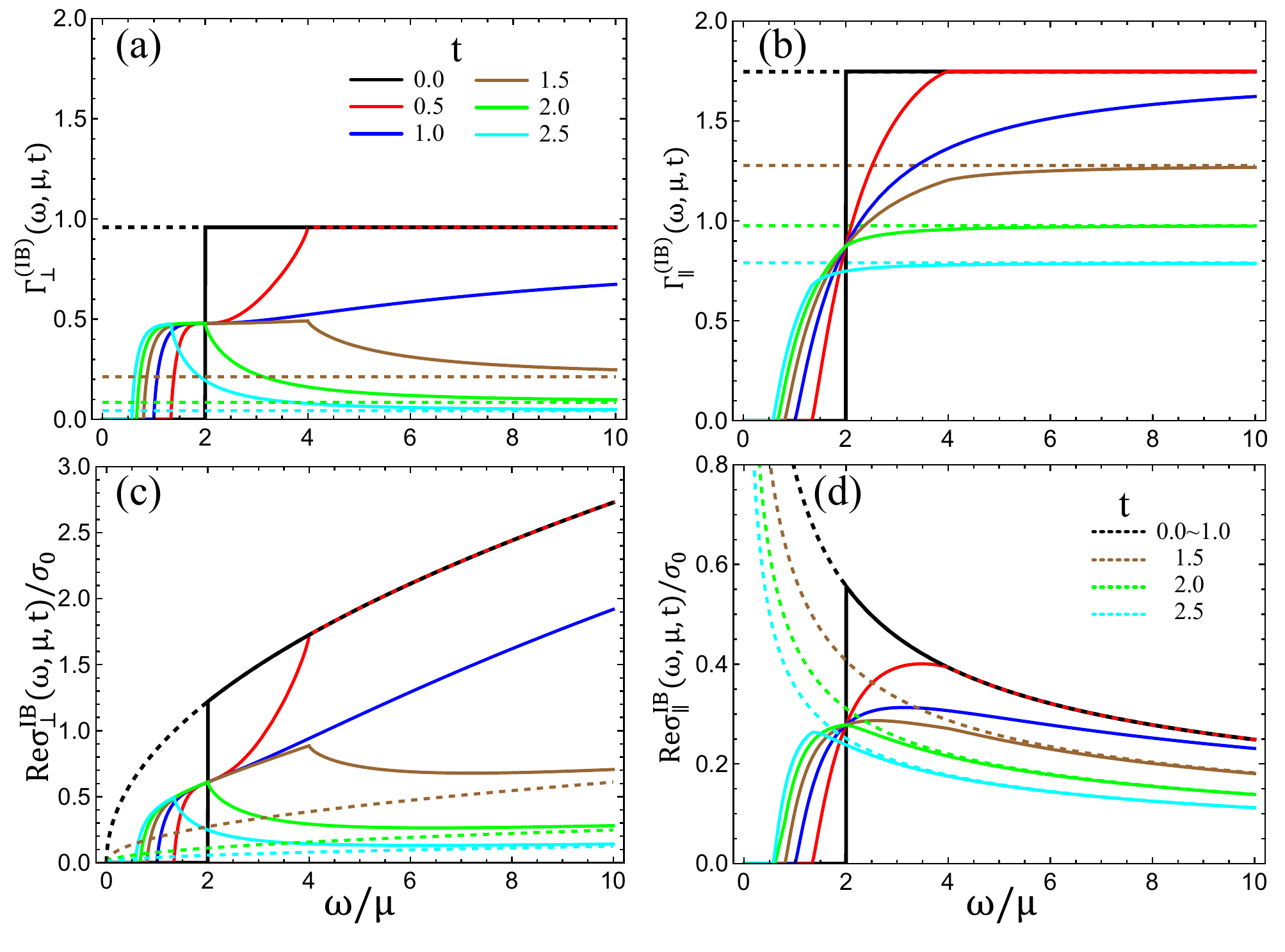}
		\caption{Dependences of $\Gamma_{\perp}^{(\mathrm{IB})}(\omega,\mu,t)$ and $\Gamma_{\parallel}^{(\mathrm{IB})}(\omega,\mu,t)$ on the photon frequecy $\omega$ are shown in panels (a) and (b), respectively. The magnitude of LOCs $\mathrm{Re}\sigma_{\chi}^{\mathrm{IB}}(\omega,\mu,t)$ is modulated by $S_{\chi}^{(\mathrm{IB})}(\omega)$ in panels (c) and (d). Dashed lines in these panels denote the case for $\mu=0$. The degeneracy factor is set to be $N_f=4$.}
		\label{LOC_IB}
	\end{figure*}

	\subsection{Unified expressions of $\Gamma_{\chi}^{(\mathrm{IB})}(\omega,\mu,t)$}
	
	To better present the following results, we introduce two useful notations
	\begin{align}
		\omega_{\pm}(t)&\equiv\frac{2\mu}{|t\mp 1|}, \\
		\zeta_{\pm}(\omega)&\equiv\frac{\omega\pm 2\mu}{\omega}\frac{\tilde{\Theta}(t)}{t},
	\end{align}
	and two auxiliary functions
	\begin{align}
		\mathcal{G}_{\perp}(x)&=\mathscr{B}(x,\frac{1}{2},\frac{3}{4})-\mathscr{B}(x,\frac{1}{2},\frac{7}{4}),\\
		\mathcal{G}_{\parallel}(x)&=\mathscr{B}(x,\frac{1}{2},\frac{5}{4}),
	\end{align}
	where
	\begin{align}
		\mathscr{B}(x;p,q)&=\frac{1}{2}\mathrm{sgn}(x)\tilde{\Theta}(x+1)
		\tilde{\Theta}(1-x)\mathrm{B}(x^2;p,q),
	\end{align}
	with $\tilde{\Theta}(x)=0$ for $x<0$ and $\tilde{\Theta}(x)=1$ for $x\ge0$, $\mathrm{sgn}(z)$ the sign function, and $\mathrm{B}(z;p,q)$ the
	incomplete Beta function. It is helpful to note that $\mathrm{B}(1;p,q)=\mathrm{B}(p,q)$ with $\mathrm{B}(p,q)$ the conventional Beta functions.
	The definition of $\mathcal{G}_{\chi}(x)$ makes it exhibit nice properties such as $\mathcal{G}_{\chi}(-x)=-\mathcal{G}_{\chi}(x)$.

	In addition, we define a more compact notation
	\begin{align}
		\tilde{\zeta}_{\pm}(\omega)=
		\begin{cases}
			+1, &\zeta_{\pm}(\omega)>+1,\\\\
			-1, &\zeta_{\pm}(\omega)<-1,\\\\
			\zeta_{\pm}(\omega), &-1\leq\zeta_{\pm}(\omega)\leq+1.
		\end{cases}
	\end{align}
	In this way, the dimensionless auxiliary function $\Gamma_{\chi}^{\mathrm{IB}}(\omega,\mu,t)$ can be expressed in a unified form as
	\begin{align}
		\Gamma_{\chi}^{(\mathrm{IB})}(\omega,\mu,t)
		=\mathcal{G}_{\chi}\left[\tilde{\zeta}_{+}(\omega)\right]
		+\mathcal{G}_{\chi}\left[\tilde{\zeta}_{-}(\omega)\right],
	\end{align}
	for both undoped and doped cases ($\mu=0$ and $\mu\neq 0$), both components ($\chi=\parallel$ and $\chi=\perp$), and all tilted phases ($t=0$, $0<t<1$, $t=1$, and $t>1$).
	
	Specifically, in the undoped case ($\mu=0$) or the asymptotic regime of large photon energy $\omega=\Omega\gg\mathrm{Max}\{\omega_{+}(t),\omega_{-}(t),2\mu\}$, we have
	\begin{align}
		\zeta_{\pm}(\omega)&=\frac{\tilde{\Theta}(t)}{t}\in
		\begin{cases}
			[+1,+\infty), & 0\leq t\leq 1,\\\\
			(0,+1), & t>1,
		\end{cases}
	\end{align}
	and hence
	\begin{align}
		\tilde{\zeta}_{\pm}(\omega)=
		\begin{cases}
			+1, &0\leq t\leq 1,\\\\
			\frac{1}{t}, &t>1.
		\end{cases}
	\end{align}
	It is convenient to introduce
	\begin{align}
		\Phi_{\chi}(t)\equiv
		\begin{cases}
			2\mathcal{G}_{\chi}(+1), & 0\leq t\leq 1,\\\\
			2\mathcal{G}_{\chi}(1/t)\in\Big(0,2\mathcal{G}_{\chi}(+1)\Big), & t>1,
		\end{cases}
	\end{align}
	which satisfies the relation
	\begin{align}
		\Gamma_{\chi}^{(\mathrm{IB})}(\omega,\mu=0,t)=\Gamma_{\chi}^{(\mathrm{IB})}(\Omega,\mu,t)=\Phi_{\chi}(t),
		\label{mu0}
	\end{align}
	for both the undoped case and the asymptotic regime. From all of the above unified expressions, it is obvious that different tilt phases of 2D tilted SDBs yield qualitatively distinct
	$\Gamma_{\chi}^{(\mathrm{IB})}(\omega,\mu,t)$ and hence $\Phi_{\chi}(t)$.
	
It is beneficial to present an intuitive picture of the interband LOCs before the formal and rigorous discussions in the next subsection. The interband LOCs are contributed by the optical transitions from the valence band to the conduction band without any change in momentum. As shown in Fig.\ref{distinct_tilt}, there are two specific optical transitions termed by two critical frequencies $\omega_{+}(t)$ and $\omega_{-}(t)$. Two remarks are in order here. Firstly, the interband LOCs are forbidden by the Pauli blocking unless $\omega \ge \omega_{-}(t)$. Basically, it predicts that the interband LOCs would be detected in the regime $\omega \ge \omega_{-}(t)$. Secondly, $\omega_+(t)$ is a critical frequency to identify the kinked behavior in the joint density of states (JDOS) \cite{PRBTan2022}, which results in the sharp feature for the interband LOCs. Further details can be found in the Subsection \ref{Subsec:JDOS}.

	\subsection{Explicit expressions of $\Gamma_{\chi}^{(\mathrm{IB})}(\omega,\mu,t)$}

	The unified expressions of $\Gamma_{\chi}^{(\mathrm{IB})}(\omega,\mu,t)$ above can be more explicitly written as follows, which are intuitively shown in Figs. \ref{LOC_IB}(a) and (b). In the untilted case ($t=0$) with $\omega_\pm(t=0)=2\mu$ \cite{PRBCarbotte2019},
	the dimensionless auxiliary function  $\Gamma_{\chi}^{(\mathrm{IB})}(\omega,\mu,t=0)=\Phi_{\chi}(0)\tilde{\Theta}(\omega-2\mu)$. For
	the type-I phase ($0<t<1$), due to the deviation between $\omega_{+}(t)$ and $\omega_{-}(t)$, the dimensionless auxiliary function $\Gamma_{\chi}^{(\mathrm{IB})}(\omega,\mu,t)$ reads
	\begin{align}
		&\hspace{-0.2cm}\Gamma_{\chi}^{(\mathrm{IB})}(\omega,\mu,0<t<1)
		\notag\\&\hspace{-0.2cm}=
		\begin{cases}
			0, & 0<\omega<\omega_{-}(t),\\\\
			\mathcal{G}_{\chi}(+1)+\mathcal{G}_{\chi}\left[\zeta_{-}(\omega)\right], & \omega_{-}(t)\leq\omega<\omega_{+}(t),\\\\
			2\mathcal{G}_{\chi}(+1), & \omega\ge\omega_{+}(t).
		\end{cases}
	\end{align}

	For the type-II phase ($t>1$), the dimensionless auxiliary function $\Gamma_{jj}^{(\mathrm{IB})}(\omega,\mu,t)$ takes
	\begin{align}
		&\hspace{-0.2cm}\Gamma_{\chi}^{(\mathrm{IB})}(\omega,\mu,t>1)
		\notag\\&\hspace{-0.2cm}=
		\begin{cases}
			0, & 0<\omega<\omega_{-}(t),\\\\
			\mathcal{G}_{\chi}(+1)+\mathcal{G}_{\chi}\left[\zeta_{-}(\omega)\right], & \omega_{-}(t)\leq\omega<\omega_{+}(t),\\\\
			\mathcal{G}_{\chi}\left[\zeta_{+}(\omega)\right]+\mathcal{G}_{\chi}\left[\zeta_{-}(\omega)\right], & \omega\ge\omega_{+}(t).
		\end{cases}
	\end{align}
	
	For the type-III phase ($t=1$), due to the relations $\omega_{-}(t)=\mu$ and $\omega_{+}(t)\to\infty$, we arrive at
	\begin{align}
		&\Gamma_{\chi}^{(\mathrm{IB})}(\omega,\mu,t=1)
		\notag\\&=
		\begin{cases}
			0, & 0<\omega<\mu,\\\\
			\mathcal{G}_{\chi}(+1)+\mathcal{G}_{\chi}\left[\zeta_{-}(\omega)\right], & \omega\ge \mu,
		\end{cases}
	\end{align}
	which can be derived from the limit of both type-I phase and type-II phase. This shows a continuity of $\Gamma_{\chi}^{(\mathrm{IB})}(\omega,\mu,t=1)$ with respect to the tilt parameter, namely,
	\begin{align}
		\lim_{ t \to1^{\pm}}\Gamma_{\chi}^{(\mathrm{IB})}(\omega,\mu,t)
		&=\Gamma_{\chi}^{(\mathrm{IB})}(\omega,\mu,t=1).
	\end{align}
	
	Four remarks on $\Gamma_{\chi}^{(\mathrm{IB})}(\omega,\mu,t)$ are in order here. First, due to the band tilting and Pauli blocking, $\Gamma_{\chi}^{(\mathrm{IB})}(\omega,\mu,t)$
	exhibits a Heaviside-like behavior, which can be divided into three different regions for both the type-I and type-II phases but two different regions for the type-III phase.
	Second, $\Gamma_{\perp}^{(\mathrm{IB})}(\omega,\mu,t)$ is generally not equal to $\Gamma_{\parallel}^{(\mathrm{IB})}(\omega,\mu,t)$. Third, $\Gamma_{\chi}^{(\mathrm{IB})}(\omega,\mu,t)$ satisfies the relation $\Gamma_{\chi}^{(\mathrm{IB})}(\omega,\mu=0,t)=\Gamma_{\chi}^{(\mathrm{IB})}(\Omega,\mu,t)=\Phi_{\chi}(t)$ for
	both the undoped case and the asymptotic regime. Fourth, for arbitrary tilt satisfying $0<t\le 2$, the fixed point of dimensionless function $\Gamma_{\chi}^{(\mathrm{IB})}(\omega=2\mu,\mu,t)$ always holds as
	\begin{align}
		\Gamma_{\chi}^{(\mathrm{IB})}(\omega=2\mu,\mu,t)=\frac{1}{2}\Phi_{\chi}(t=0),
		\label{converge}
	\end{align}
	where $\Phi_{\chi}(t=0)$ satisfies $\Phi_{\perp}(t=0)\neq\Phi_{\parallel}(t=0)$. Anyhow, Eq.(\ref{converge}) accounts for the fixed point for arbitrary tilt $t$ satisfying $t\in (0,2]$ in $\mathrm{Re}\sigma_{\chi}^{\mathrm{IB}}(\omega,\mu,t)$ and in $\sqrt{\mathrm{Re}\sigma_{\perp}^{\mathrm{IB}}(\omega,\mu,t)\times\mathrm{Re}\sigma_{\parallel}^{\mathrm{IB}}(\omega,\mu,t)}$, which is also reported in 2D tilted Dirac bands \cite{arXivHou2022}. And, we would unveil its physical origin in the next subsection.

\begin{figure*}[htbp]
		\includegraphics[width=16cm]{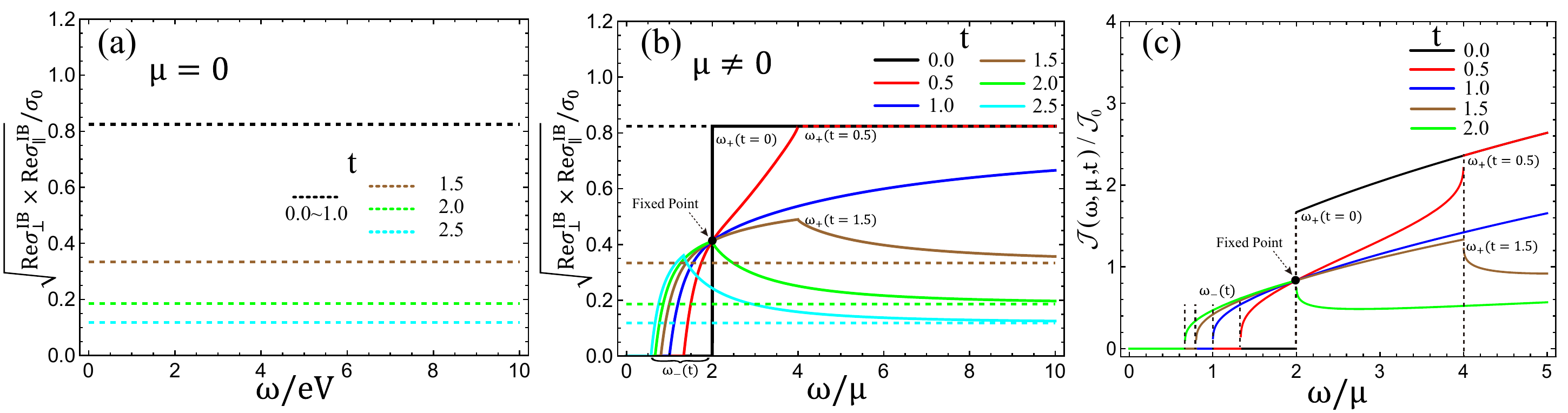}
		\caption{Dependences of $\sqrt{\mathrm{Re}\sigma_{\perp}(\omega,\mu,t)\times\mathrm{Re}\sigma_{\parallel}(\omega,\mu,t)}$ and the JDOS on the photon frequency $\omega$. The dashed lines denote the undoped case ($\mu=0$) in (a) and the asymptotic background value in (b), whereas the solid lines in (b) yield the doped case ($\mu\neq0$). The JDOS for distinct tilts are plotted in panel (c). The degeneracy factor is set to be $N_f=4$. In panels (a) and (b), the arguments $(\omega,\mu,t)$ are omitted in the product $\sqrt{\mathrm{Re}\sigma_{\perp}(\omega,\mu,t)\times\mathrm{Re}\sigma_{\parallel}(\omega,\mu,t)}$.}
		\label{figasymp}
\end{figure*}

	\subsection{Results for $\mathrm{Re}\sigma_{\chi}^{\mathrm{IB}}(\omega,\mu,t)$}
	
	With the help of the analytical expressions of $S_{\chi}^{(\mathrm{IB})}(\omega)$ and $\Gamma_{\chi}^{(\mathrm{IB})}(\omega,\mu,t)$, we plot $\mathrm{Re}\sigma_{\chi}^{\mathrm{IB}}(\omega,\mu,t)$ in Figs.\ref{LOC_IB}(c) and (d). As expected in the untilted case ($t=0$) \cite{PRBCarbotte2019}, $\mathrm{Re}\sigma_{\chi}^{\mathrm{IB}}(\omega,\mu,t)$ (the solid lines) are scaled by the dimensional auxiliary function $S_{\chi}^{(\mathrm{IB})}(\omega)$ (the black dashed lines) as $\sqrt{\omega}$ in Fig.\ref{LOC_IB}(c) and $1/\sqrt{\omega}$ in Fig.\ref{LOC_IB}(d). The dashed lines display the asymptotic behaviors in both the regime of large photon energy $\omega=\Omega\gg\omega_+$ and the undoped case $\mu=0$.
	
	Analytically, it is evident that for both
	untilted phase \cite{PRBCarbotte2019} and type-I phase, the asymptotic background value takes
	\begin{align}
		\mathrm{Re}\sigma_{\chi}^{\mathrm{asymp}}(\omega,\mu,0\leq t<1)
		&=N_{f}S_{\chi}^{(\mathrm{IB})}(\Omega)2\mathcal{G}_{\chi}(+1),
	\end{align}
	where
	\begin{align}
		\mathcal{G}_{\perp}(+1)&=\frac{\mathrm{B}(\frac{1}{2},\frac{3}{4})-\mathrm{B}(\frac{1}{2},\frac{7}{4})}{2},\\
		\mathcal{G}_{\parallel}(+1)&=\frac{\mathrm{B}(\frac{1}{2},\frac{5}{4})}{2},
	\end{align}
	with $\mathrm{B}(p,q)=\mathrm{B}(+1;p,q)$ the conventional Beta function. Accordingly, the product of asymptotic background values takes
	\begin{align}
		&\mathrm{Re}\sigma_{\perp}^{\mathrm{asymp}}(\omega,\mu,0\leq t<1)\times
		\mathrm{Re}\sigma_{\parallel}^{\mathrm{asymp}}(\omega,\mu,0\leq t<1)
		\nonumber\\&
		=\mathcal{G}_{\perp}(+1)\mathcal{G}_{\parallel}(+1)\frac{N_{f}^2\sigma_0^2}{\pi^2}
		=\frac{32}{15\pi}\frac{N_{f}^2\sigma_0^2}{16}.
		\label{asyp1}
	\end{align}
	
	For the type-II phase, due to $\zeta_{\pm}(\Omega)=1/t$, we have the asymptotic background value
	\begin{align}
		\mathrm{Re}\sigma_{\chi}^{\mathrm{asymp}}(\omega,\mu,t>1)
		&=N_{f}S_{\chi}^{(\mathrm{IB})}(\Omega)2\mathcal{G}_{\chi}(1/t),
	\end{align}
	where
	\begin{align}
		\mathcal{G}_{\perp}(1/t)&=\frac{\mathrm{B}(\frac{1}{t};\frac{1}{2},\frac{3}{4})-\mathrm{B}(\frac{1}{t};\frac{1}{2},\frac{7}{4})}{2},\\
		\mathcal{G}_{\parallel}(1/t)&=\frac{\mathrm{B}(\frac{1}{t};\frac{1}{2},\frac{5}{4})}{2}.
	\end{align}
	The product of asymptotic background values can accordingly be written as
	\begin{align}
		&\mathrm{Re}\sigma_{\perp}^{\mathrm{asymp}}(\omega,\mu,t>1)\times
		\mathrm{Re}\sigma_{\parallel}^{\mathrm{asymp}}(\omega,\mu,t>1)
		\nonumber\\&
		=\mathcal{G}_{\perp}(1/t)\mathcal{G}_{\parallel}(1/t)\frac{N_{f}^2\sigma_0^2}{\pi^2},
		\label{asyp1}
	\end{align}
	which is tilt-dependent due to $\mathcal{G}_{\perp}(1/t)\mathcal{G}_{\parallel}(1/t)$.

	For the type-III phase, we have the asymptotic background value
	\begin{align}
		\mathrm{Re}\sigma_{\chi}^{\mathrm{asymp}}(\omega,\mu,t=1)
		&=N_{f}S_{\chi}^{(\mathrm{IB})}(\Omega)2\mathcal{G}_{\chi}(+1),
	\end{align}
	and their product
	\begin{align}
		&\mathrm{Re}\sigma_{\perp}^{\mathrm{asymp}}(\omega,\mu,t=1)\times
		\mathrm{Re}\sigma_{\parallel}^{\mathrm{asymp}}(\omega,\mu,t=1)
		\nonumber\\&
		=\mathcal{G}_{\perp}(+1)\mathcal{G}_{\parallel}(+1)\frac{N_{f}^2\sigma_0^2}{\pi^2}
		=\frac{32}{15\pi}\frac{N_{f}^2\sigma_0^2}{16},
	\end{align}
	which can also be obtained by taking the limit of both type-I phase and type-II phase.

For distinct tilt parameters, the dependences of $\sqrt{\mathrm{Re}\sigma_{\perp}(\omega,\mu,t)\times\mathrm{Re}\sigma_{\parallel}(\omega,\mu,t)}$ on the photon frequency $\omega$ are depicted in Fig.\ref{figasymp}.
	
	Interestingly, for $0 < t\leq 2$, the LOCs in 2D tilted SDBs take the forms
	\begin{align}
		&\mathrm{Re}\sigma_{\perp}^{\mathrm{IB}}(\omega=2\mu,\mu,0 < t\leq 2)
		\nonumber\\&
		=N_{f}\sigma_0\frac{\sqrt{a\mu}}{2\pi v_F}\left[\mathrm{B}(\frac{1}{2},\frac{3}{4})-\mathrm{B}(\frac{1}{2},\frac{7}{4})\right],
	\end{align}
	and
	\begin{align}
		&\mathrm{Re}\sigma_{\parallel}^{\mathrm{IB}}(\omega=2\mu,\mu,0 < t\leq 2)\nonumber\\
		&=N_{f}\sigma_0\frac{ v_F}{2\pi\sqrt{a\mu}}\frac{1}{4}\mathrm{B}(\frac{1}{2},\frac{5}{4}).
	\end{align}
After utilizing the relation $\left[\mathrm{B}(\frac{1}{2},\frac{3}{4})-\mathrm{B}(\frac{1}{2},\frac{7}{4})\right]
\mathrm{B}(\frac{1}{2},\frac{5}{4})
=\frac{8\pi}{15}$, one obtains the product 
\begin{align}
&\mathrm{Re}\sigma_{\perp}^{\mathrm{IB}}(2\mu,\mu,0 < t\leq 2)\times
\mathrm{Re}\sigma_{\parallel}^{\mathrm{IB}}(2\mu,\mu,0 < t\leq 2)
\nonumber\\&
=\frac{8}{15\pi}\frac{N_{f}^2\sigma_0^2}{16},
\end{align}
which is a constant independent of $\mu$, $a$, and $v_F$.

\begin{widetext}	
	
\subsection{Intuitive understandings from JDOS 
\label{Subsec:JDOS}}

The interband LOCs at $\omega$ are generally determined by the interband optical transition. The number of states involved in the transition can be calculated by the JDOS via
\begin{align}
\mathcal{J}(\omega,\mu,t)&=N_f\int \frac{d^2 \boldsymbol{k}}{(2\pi)^2}\left\{f\left[\varepsilon_{+}^{-}(k_x,k_y)\right] -f\left[\varepsilon_{+}^{+}(k_x,k_y)\right]\right\}
	\delta\left[\omega+\varepsilon_{+}^{-}(k_x,k_y)-\varepsilon_{+}^{+}(k_x,k_y)\right],
	\label{eqnJDOS}
\end{align}
where $\left\{f\left[\varepsilon_{+}^{-}(k_x,k_y)\right] -f\left[\varepsilon_{+}^{+}(k_x,k_y)\right]\right\}$ and $\delta\left[\omega+\varepsilon_{+}^{-}(k_x,k_y)-\varepsilon_{+}^{+}(k_x,k_y)\right]$ accounts for the Pauli exclusion principle and energy conservation of optical transition, respectively. At zero temperature, the JDOS for $0<t<1$ reads 
\begin{align}
\frac{\mathcal{J}(\omega,\mu,t)}{\mathcal{J}_0}&=\frac{N_f}{4\pi}\sqrt{\frac{\omega}{2\mu}}\times
	\begin{cases}
		0,&0\leq\omega<\omega_{-}(t),\\\\
\left[\mathscr{B}\left(1,\frac{1}{2},\frac{1}{4}\right)
+\mathscr{B}\left(\xi_{-},\frac{1}{2},\frac{1}{4}\right)\right],&\omega_{-}(t)\leq\omega<\omega_{+}(t),\\\\
		2\mathscr{B}\left(1,\frac{1}{2},\frac{1}{4}\right),&\omega\ge\omega_{+}(t),
	\end{cases}
\end{align}
while the JDOS for $t>1$ takes 
\begin{align}
\frac{\mathcal{J}(\omega,\mu,t)}{\mathcal{J}_0}&=\frac{N_f}{4\pi}\sqrt{\frac{\omega}{2\mu}}\times
\begin{cases}
		0,&0\leq\omega<\omega_{-}(t),\\\\ 
\left[\mathscr{B}\left(1,\frac{1}{2},\frac{1}{4}\right)
+\mathscr{B}\left(\xi_{-},\frac{1}{2},\frac{1}{4}\right)\right],&\omega_{-}(t)\leq\omega<\omega_{+}(t),\\\\ \left[\mathscr{B}\left(\xi_{-},\frac{1}{2},\frac{1}{4}\right)
+\mathscr{B}\left(\xi_{+},\frac{1}{2},\frac{1}{4}\right)\right],&\omega\ge\omega_{+}(t),
	\end{cases}
\end{align}
where $\mathcal{J}_0=\frac{\sqrt{\mu}}{2\pi\sqrt{a} v_F}$.

\begin{figure*}[htbp]
	\includegraphics[width=12cm]{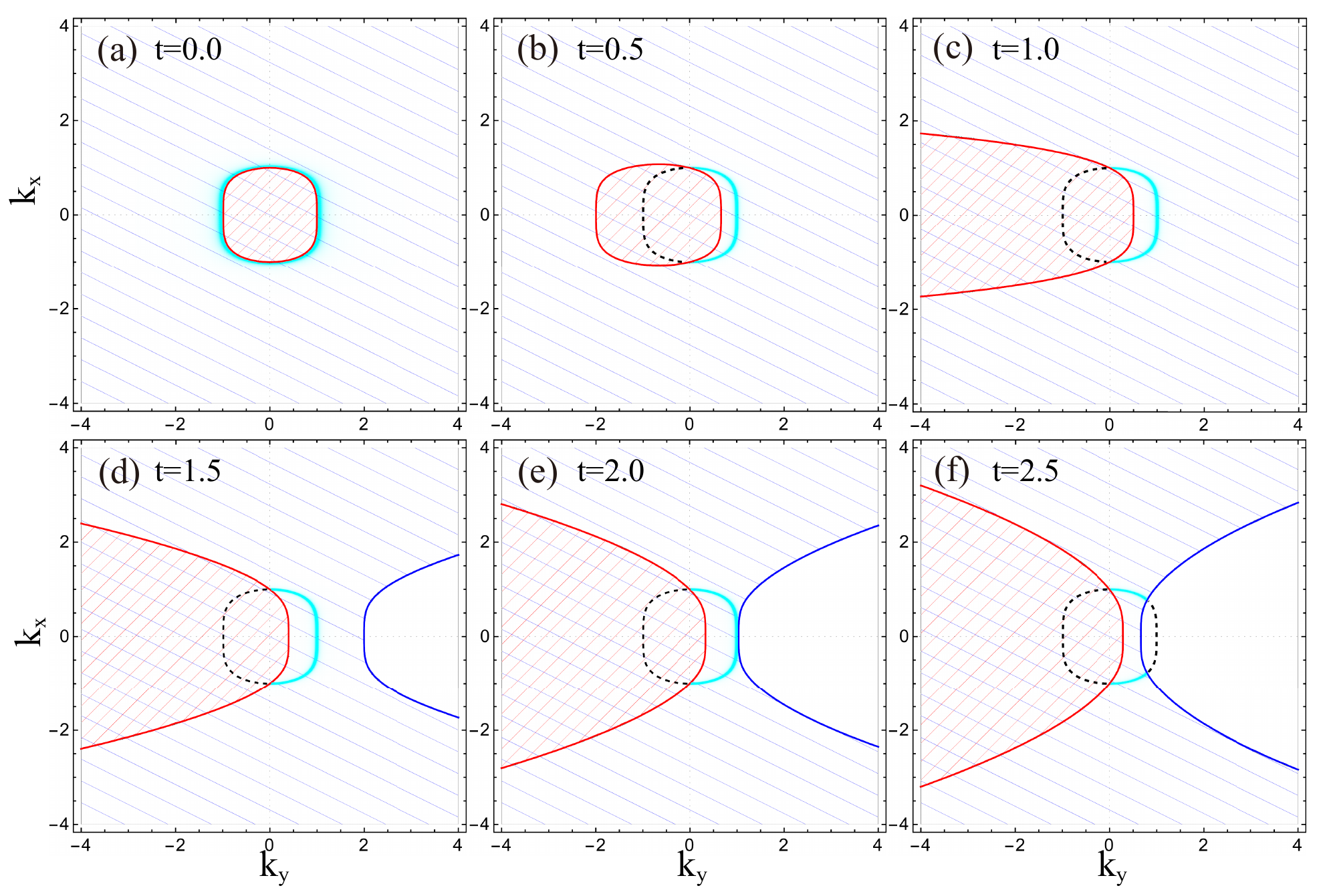}
	\caption{In 2D tilted SDBs, the states involved in the interband optical transition with $\omega=2\mu$ are gathered around a cyan stripe in panels (a) - (f). The Fermi surfaces are denoted by red and blue lines. The location of states occupying below the Fermi surface is meshed with color slashes.}
	\label{states}
\end{figure*}

\end{widetext}

As shown in Fig.\ref{figasymp}(c), the non-trivial JDOS $\mathcal{J}(\omega,\mu,t)$ is pumped out from $\omega=\omega_{-}(t)$, as no state participates in the interband optical transition unless $\omega \ge \omega_{-}(t)$. It predicts that the frequency of interband LOCs in Fig.\ref{figasymp}(b) starts from $\omega=\omega_{-}(t)$. The discontinuity in the first derivative of JDOS at $\omega=\omega_{+}(t)$ results in the kinked feature of $\mathcal{J}(\omega,\mu,t)$ therein, which is responsible for the sharp feature of the interband LOCs at $\omega=\omega_{+}(t)$, as shown in Fig.\ref{figasymp}(b).

To intuitively visualize the JDOS at $\omega=2\mu$, we identify the states involved in the interband optical transition by cyan stripes as shown in Fig.\ref{states}, where the red and blue lines denote the contours of Fermi surface. As apparently shown in Figs.\ref{states}(b)-(e), for $0<t\le2$, the states involved in the interband optical transition with $\varepsilon_\kappa^{+}(k_x,k_y)-\varepsilon_\kappa^{-}(k_x,k_y)=2\mu$ gather around the same cyan strip. These states are exactly counted by
\begin{align}
\mathcal{J}(\omega=2\mu,\mu,0<t\le2)=\frac{N_f \mathcal{J}_0}{8\pi}\mathrm{B}(\frac{1}{2},\frac{1}{4}),
\end{align}
which precisely give a fixed point for arbitrary tilt $t$ satisfying $t\in (0,2]$ in Fig.\ref{states}(c). As a result, the corresponding fixed points of $\Gamma_{\chi}^{(\mathrm{IB})}(\omega=2\mu,\mu,t)$, $\mathrm{Re}\sigma_{\chi}^{\mathrm{IB}}(\omega,\mu,t)$, and $\sqrt{\mathrm{Re}\sigma_{\perp}^{\mathrm{IB}}(\omega,\mu,t)\times\mathrm{Re}\sigma_{\parallel}^{\mathrm{IB}}(\omega,\mu,t)}$ appear in Fig.\ref{LOC_IB} and Fig.\ref{figasymp}(b), regardless of the tilt parameter $t$ for $0<t\le2$. 

Interestingly, the number of states participating in the interband optical transition for tilted cases ($0<t\le 2$) is exactly half of that for untilted case ($t=0$), as shown in Fig.\ref{states}. This result can be equivalently formulated by 
\begin{align}
\mathcal{J}(\omega=2\mu,\mu,0<t\le2)=\frac{\mathcal{J}(\omega=2\mu,\mu,t=0)}{2},
\end{align}
which is also shown in Fig.\ref{figasymp}(c). 

Furthermore, as shown in Fig.\ref{figasymp}(c) and Fig.\ref{states}(f), the amount of states participating in the optical transition starts to decrease with the tilt parameter increasing from $t=2$. Thus, $\Gamma_{\chi}^{(\mathrm{IB})}(\omega=2\mu,\mu,t)$ and $\mathrm{Re}\sigma_{\chi}^{\mathrm{IB}}(\omega=2\mu,\mu,t)$ for $t>2$ are always smaller than their counterparts for $0<t\le 2$. A similar analysis for 2D tilted Dirac bands has been performed more detailedly in a recent work by the authors of this manuscript \cite{arXivHou2022}. Together with the analysis therein, the robust behavior of fixed point is universal in spite of different geometric structure of Fermi surface. In this sense, we report a universal and robust behavior of fixed point at $\omega=2\mu$.

\begin{widetext}	
	
\subsection{Angular dependence of LOCs}

	\begin{figure*}[htp]
		\includegraphics[width=12cm]{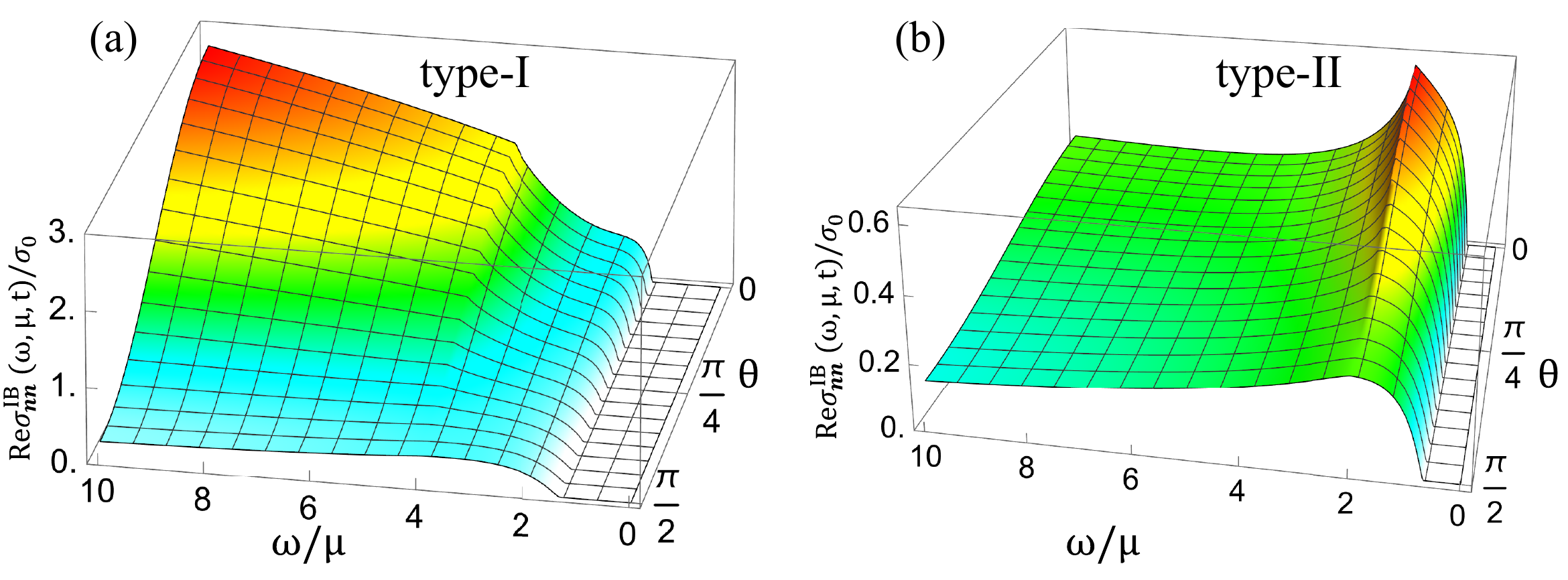}
		\caption{Angular dependence of interband LOCs in 2D tilted SDBs. The degeneracy factor is taken to be $N_f=4$.}
		\label{angular}
	\end{figure*}
	
We are also interested in the angular dependence of LOCs, which can be extracted from
\begin{align}
\sigma_{nn}^{\kappa}(\omega,\mu,t)&=\frac{i}{\omega}
\frac{1}{\beta}\sum_{\Omega_m}\int_{-\infty}^{+\infty}\frac{dk_x}{2\pi}\int_{-\infty}^{+\infty}\frac{dk_y}{2\pi}
\mathrm{Tr}\left[\hat{J}_{n}^{\kappa}G_{\kappa}(k_x,k_y,i\Omega_m,\mu)
\hat{J}_{n}^{\kappa}G_{\kappa}(k_x,k_y,i\Omega_m+\omega+i\eta,\mu)\right],
\end{align}
where $\hat{J}_{n}^{\kappa}=\hat{J}_{x}^{\kappa}\mathrm{cos}\theta+\hat{J}_{y}^{\kappa}\mathrm{sin}\theta$ with $\theta$ the angle between the basis vector $\boldsymbol{e}_x$ and an arbitrary basis vector $\boldsymbol{n}=(\mathrm{cos}\theta,\mathrm{sin}\theta)$ in 2D. Hence, the LOCs along arbitrary direction reads,
	\begin{align}
		\mathrm{Re}\sigma_{nn}^{\kappa}(\omega,\mu,t)&=
		\mathrm{Re}\sigma_{xx}^{\kappa}(\omega,\mu,t)\mathrm{cos}^2\theta
+\mathrm{Re}\sigma_{yy}^{\kappa}(\omega,\mu,t)\mathrm{sin}^2\theta,
	\end{align}
	where we have used  the property of vanishing transverse optical conductivities $\mathrm{Re}\sigma_{xy}^{\kappa}(\omega,\mu,t)$ and $\mathrm{Re}\sigma_{yx}^{\kappa}(\omega,\mu,t)$. 

As explicitly shown in Fig.\ref{angular}, the angular dependence of interband LOCs exhibits a strong anisotropy. In addition, it differs significantly from the angular dependence of interband LOCs for 2D tilted Dirac bands \cite{arXivHou2022}.
	
\end{widetext}

	\section{INTRABAND CONDUCTIVTY \label{Sec:Intraband conductivity}}
	
	In this section, we turn to the intraband (Drude) conductivity, which is contributed by the intraband transition around the Fermi surface by utilizing Eq.(9),
	where the derivative of the Fermi distribution function can be replaced by $\delta[\mu -\varepsilon_\kappa^\lambda(k_x,k_y)]$ at zero temperature. Similarly,
	the intraband conductivity can be recast as
	\begin{align}
		&\mathrm{Re}\sigma_{jj}^{\mathrm{D}}(\omega,\mu,t)
\nonumber\\&
=\mathrm{Re}\sigma_{\perp}^{\mathrm{D}}(\omega,\mu,t)\delta_{jx}+
		\mathrm{Re}\sigma_{\parallel}^{\mathrm{D}}(\omega,\mu,t)\delta_{jy},
	\end{align}
	where
	\begin{align}
		\mathrm{Re}\sigma_{\chi}^{\mathrm{D}}(\omega,\mu,t)
		=&N_{f}S_{\chi}^{(\mathrm{D})}(\mu)\Gamma_{\chi}^{(\mathrm{D})}(\Lambda,\mu,t)\delta(\omega).
		\label{LOCsp2}
	\end{align}
	The essential information of intraband conductivity is encoded into two auxiliary functions $S_{\chi}^{(\mathrm{D})}(\mu)$ and $\Gamma_{\chi}^{(\mathrm{D})}(\Lambda,\mu,t)$, whose detailed calculations are found in Appendix  \ref{App4}. First, the dimensional auxiliary function $S_{\chi}^{(\mathrm{D})}(\mu)$ can be explicitly written as
	\begin{align}
		S_{\perp}^{(\mathrm{D})}(\mu)&=\frac{\sigma_{0}\mu}{2\pi}\frac{\sqrt{a\mu}}{\upsilon_F},\\
		S_{\parallel}^{(\mathrm{D})}(\mu)&=\frac{\sigma_{0}\mu}{2\pi}\frac{\upsilon_F}{\sqrt{a\mu}},
	\end{align}
	which satisfy the relation $S_{\perp}^{(\mathrm{D})}(\mu)\times S_{\parallel}^{(\mathrm{D})}(\mu)=\frac{\sigma_{0}^2\mu^2}{4\pi^2}$. It is stressed that $S_{\chi}^{(\mathrm{D})}(\mu)$ is independent of tilt parameter, hence agrees with the results in 2D untilted SDBs \cite{PRBCarbotte2019}. Second, the explicit expressions of dimensionless auxiliary function $\Gamma_{\chi}^{(\mathrm{D})}(\Lambda,\mu,t)$ depends on $\Lambda$ and $\mu$ in terms of $\tilde{\Lambda}=\Lambda/\mu$, where the momentum cutoff $\Lambda$ was introduced to prevent divergence from calculation, with $\upsilon_F k_y<\Lambda$. 
	In the untilted case ($t=0$), the dimensionless auxiliary functions take
	\begin{align}
		\begin{cases}
			\Gamma_{\perp}^{(\mathrm{D})}(\tilde{\Lambda},t=0)=\frac{48\sqrt{2}}{5\sqrt{\pi}}
			\left[\mathrm{B}\left(\frac{3}{4},\frac{3}{4}\right)\right]^2,\\\\
			\Gamma_{\parallel}^{(\mathrm{D})}(\tilde{\Lambda},t=0)=\frac{2\sqrt{2}}{3\sqrt{\pi}}
			\left[\mathrm{B}\left(\frac{1}{4},\frac{1}{4}\right)\right]^2,
		\end{cases}
	\end{align}
	which are independent of $\tilde{\Lambda}$ and restore previous result for untilted case \cite{PRBCarbotte2019}. For sake of clarity, some detailed definitions for auxiliary function we present below can be found in Appendix \ref{App4}.
	Turn to the type-I phase ($0<t<1$),
	\begin{align}
		&\Gamma_{\perp}^{(\mathrm{D})}(\Lambda,\mu,0<t<1)
		\nonumber\\&=
		\mathrm{B}\left(\frac{7}{4},\frac{7}{4}\right)\mathscr{X}\left(+;\frac{7}{4};1,1;\frac{7}{2};t X_u^+(t),0\right),
	\end{align}
	and
	\begin{align}
		&\Gamma_{\parallel}^{(\mathrm{D})}(\Lambda,\mu,0<t<1)
		\nonumber\\&=
		\mathrm{B}\left(\frac{1}{4},\frac{1}{4}\right)\mathscr{Y}\left(+;\frac{1}{4} ;1,1;\frac{1}{2};t X_u^+(t),0\right),
	\end{align}
	where
	\begin{align}
		X_u^\lambda(t)=\frac{\tilde{\omega}_{-}+\tilde{\omega}_{+}}{1+t\lambda \tilde{\omega}_{\lambda}}.
	\end{align}
	where $\tilde{\omega}_{\pm}=\omega_{\pm}(t)/\mu$ and $\lambda=\pm$. Obviously, $\Gamma_{\perp}^{(\mathrm{D})}(\tilde{\Lambda},0<t<1)$ and $\Gamma_{\parallel}^{(\mathrm{D})}(\tilde{\Lambda},0<t<1)$ are always convergent as $\tilde{\Lambda}\to+\infty$, and automatically give rise to the untilted counterpart $\Gamma_{\perp}^{(\mathrm{D})}(\Lambda,\mu,t=0)$ and $\Gamma_{\parallel}^{(\mathrm{D})}(\Lambda,\mu,t=0)$ by taking the limit $t\to0^{+}$.
	
	For the type-II phase, the dimensionless functions
	\begin{align}
		&\Gamma_{\perp}^{(\mathrm{D})}(\Lambda,\mu,t>1)	
		=\sum_{\lambda=\pm}\mathrm{B}\left(\frac{7}{4},1\right)\lambda\left[Y_o^{\lambda}(t,\tilde{\Lambda})\right]^{7/4}
		\nonumber\\&\times
		\mathscr{X}\left(\lambda;\frac{7}{4};1,-\frac{3}{4};\frac{11}{4};-tX_o^{\lambda}(t,\tilde{\Lambda}),-Y_o^{\lambda}(t,\tilde{\Lambda})\right),
	\end{align}
	and
	\begin{align}
		&\Gamma_{\parallel}^{(\mathrm{D})}(\Lambda,\mu,t>1)
		=\sum_{\lambda=\pm}\mathrm{B}\left(\frac{1}{4},1\right)\lambda\left[Y_o^{\lambda}(t,\tilde{\Lambda})\right]^{1/4}
		\nonumber\\&\times
		\mathscr{Y}\left(\lambda;\frac{1}{4};1,\frac{3}{4};\frac{7}{4};-tX_o^{\lambda}(t,\tilde{\Lambda}),-Y_o^{\lambda}(t,\tilde{\Lambda})\right),
	\end{align}
	where $X_o(\lambda,t)$ and $Y_o(\lambda,t)$ are defined as
	\begin{align}
		X_o^\lambda(t,\tilde{\Lambda})&=\frac{\tilde{\Lambda}+\lambda\tilde{\omega}_{-\lambda}}{\lambda-t\lambda\tilde{\omega}_{-\lambda}},\\
		Y_o^\lambda(t,\tilde{\Lambda})&=\frac{\tilde{\Lambda}+\lambda\tilde{\omega}_{-\lambda}}{\tilde{\omega}_{+}-\tilde{\omega}_{-}},
	\end{align}
	with $\lambda=\pm$. For the type-II phase, $\Gamma_{\chi}^{(\mathrm{D})}(\Lambda,\mu,t>1)$ are always divergent as $\tilde{\Lambda}$ approaches to infinity.
	
	For the type-III phase ($t=1$),
	\begin{align}
		&\Gamma_{\perp}^{(\mathrm{D})}(\Lambda,\mu,t=1)=8\left(1+2\tilde{\Lambda}\right)^{3/4}
		\frac{1+2\tilde{\Lambda}}{2+2\tilde{\Lambda}}
		\nonumber\\&\hspace{1.5cm}\times
		F_{1}\left(1;-\frac{3}{4},1;2;1,\frac{1+2\tilde{\Lambda}}{2+2\tilde{\Lambda}}\right),
	\end{align}
	and
	\begin{align}
		&\Gamma_{\parallel}^{(\mathrm{D})}(\Lambda,\mu,t=1)
		=2\left(1+2\tilde{\Lambda}\right)^{-3/4} \frac{1+2\tilde{\Lambda}}{2+2\tilde{\Lambda}}
		\nonumber\\&\hspace{1.5cm}\times
		F_{1}\left(1;\frac{3}{4},1;2;1,\frac{1+2\tilde{\Lambda}}{2+2\tilde{\Lambda}}\right).
	\end{align}
	
	Keeping up to the finite terms when $\tilde{\Lambda}\to\infty$, one has
\begin{align}
\Gamma_{\parallel}^{(\mathrm{D})}(\Lambda,\mu,t=1)&=2\sqrt{2}\pi+\mathcal{O}\left(\tilde{\Lambda}^{-3/4}\right),\\
\Gamma_{\perp}^{(\mathrm{D})}(\Lambda,\mu,t=1)&=\frac{32}{3}\tilde{\Lambda}^{3/4}-8\sqrt{2}\pi+\mathcal{O}\left(\tilde{\Lambda}^{-1/4}\right),
\end{align}
which indicate that $\Gamma_{\parallel}^{(\mathrm{D})}(\Lambda,\mu,t=1)=2\sqrt{2}\pi$ is convergent but $\Gamma_{\perp}^{(\mathrm{D})}(\Lambda,\mu,t=1)$ is divergent.

	\section{Discussion and Summary\label{Sec: Discussion and Summary}}
	
	The qualitative difference between interband LOCs
	$\mathrm{Re}\sigma_{\perp}^{\mathrm{IB}}(\omega,\mu,t)$ and $\mathrm{Re}\sigma_{\parallel}^{\mathrm{IB}}(\omega,\mu,t)$ is dictated by both the
	power-law scaling of $\omega$ in $S_{\chi}^{(\mathrm{IB})}(\omega)$ and the dimensionless auxiliary function of $\omega/\mu$ in $\Gamma_{\chi}^{(\mathrm{IB})}(\omega,\mu,t)$. Similarly, the essential difference between intraband LOCs
	$\mathrm{Re}\sigma_{\perp}^{\mathrm{D}}(\omega,\mu,t)$ and $\mathrm{Re}\sigma_{\parallel}^{\mathrm{D}}(\omega,\mu,t)$ is determined by both
	the power-law scaling of $\mu$ in $S_{\chi}^{(\mathrm{D})}(\mu)$ and the dimensionless auxiliary function of $\Lambda/\mu$ in $\Gamma_{\chi}^{(\mathrm{D})}(\Lambda,\mu,t)$. It is noted
	that distinct power-law scalings between $S_{\perp}^{(\mathrm{IB})}(\omega)$ and
	$S_{\parallel}^{(\mathrm{IB})}(\omega)$, or between $S_{\perp}^{(\mathrm{D})}(\mu)$ and $S_{\parallel}^{(\mathrm{D})}(\mu)$, are originated
	from the semi-Dirac dispersion (quadratic dispersion in
	$k_x$ but linear in $k_y$), but different behaviors between
	$\Gamma_{\perp}^{(\mathrm{IB})}(\omega,\mu,t)$ and $\Gamma_{\parallel}^{(\mathrm{IB})}(\omega,\mu,t)$, or between $\Gamma_{\perp}^{(\mathrm{D})}(\Lambda,\mu,t)$ and $\Gamma_{\parallel}^{(\mathrm{D})}(\Lambda,\mu,t)$, are resulted from both the 2D SDBs and band tilting.
	
	\begin{table}[htbp]
		\begin{tabular*}
			{\columnwidth}{@{\extracolsep{\fill}} c | c c}
			\hline
			\hline
			$ $  &$\mathrm{tilted~semi}$-$\mathrm{Dirac~bands}$ &$\mathrm{tilted~Dirac~bands}$
			\tabularnewline
			\hline
			$S_{\perp}^{(\mathrm{IB})}(\omega)$  &$\frac{\sigma_0}{2\pi}\frac{\sqrt{2a\omega}}{v_F}$ &$\sigma_{0}$
			\tabularnewline
			\hline
			$S_{\parallel}^{(\mathrm{IB})}(\omega)$   &$\frac{\sigma_0}{2\pi}\frac{v_F}{\sqrt{2a\omega}}$ &$\sigma_{0}$
			\tabularnewline
			\hline
			$S_{\perp}^{(\mathrm{D})}(\mu)$  &$\frac{\sigma_0\mu}{2\pi}\frac{\sqrt{a \mu}}{v_F}$ &$4\sigma_{0}\mu$
			\tabularnewline
			\hline
			$S_{\parallel}^{(\mathrm{D})}(\mu)$   &$\frac{\sigma_0\mu}{2\pi}\frac{v_F}{\sqrt{a \mu}}$ &$4\sigma_{0}\mu$
			\tabularnewline
			\hline
			\hline
		\end{tabular*}
		\caption{Power-law scalings of $S_{\chi}^{(\mathrm{IB})}(\omega)$ and $S_{\chi}^{(\mathrm{D})}(\mu)$ in 2D tilted semi-Dirac bands and 2D tilted Dirac bands.}
		\label{Tab1}
	\end{table}

	To highlight these qualitative characteristics, we further compare the LOCs for 2D tilted SDBs with that
	for 2D tilted Dirac bands and 2D untilted SDBs. First,
	we list the explicit expressions of $S_{\chi}^{(\mathrm{IB})}(\omega)$ and $S_{\chi}^{(\mathrm{D})}(\mu)$
	for 2D tilted SDBs and 2D tilted Dirac bands in Table \ref{Tab1}. Different from isotropic dimensional auxiliary function $S_{\chi}^{(\mathrm{IB})}(\omega)$
	in 2D tilted Dirac bands \cite{PRBTan2022}, the anisotropic
	dimensional auxiliary functions $S_{\perp}^{(\mathrm{IB})}(\omega)$ and $S_{\parallel}^{(\mathrm{IB})}(\omega)$ in 2D tilted Dirac bands depend on $\omega$ in a different manner. Similarly, $S_{\chi}^{(\mathrm{IB})}(\omega)$ in 2D tilted Dirac bands \cite{arXivHou2022}
	is also isotropic, whereas $S_{\perp}^{(\mathrm{D})}(\mu)$ and $S_{\parallel}^{(\mathrm{D})}(\mu)$ in 2D
	tilted SDBs show a strong anisotropy in $\mu$. Second,
	$S_{\chi}^{(\mathrm{IB})}(\omega)$ and $S_{\chi}^{(\mathrm{D})}(\mu)$ in 2D tilted SDBs are free of tilt parameter, hence are the same to that in 2D untilted SDBs \cite{PRXRoy2018,PRBCarbotte2019}. Third, $\Gamma_{\chi}^{(\mathrm{IB})}(\omega,\mu,t)$ generally behaves as a step-like function, similar to that in 2D tilted
	Dirac bands \cite{PRBTan2022}. The tilt-dependent behaviors of LOCs
	can qualitatively distinguish 2D tilted SDBs from 2D untilted
	SDBs, but show similarities in the impact of band tilting on the LOCs between 2D tilted SDBs and 2D tilted
	Dirac bands. In particular, the robust behavior of fixed
	point is similar to that in 2D tilted Dirac bands where
	$\Gamma_{\chi}^{(\mathrm{IB})}(\omega=2\mu,\mu,t)=\Phi_{\chi}(t=0)/2$ \cite{arXivHou2022}, but $\Phi_{\chi}(t=0)$
	therein satisfies $\Phi_{\perp}(t=0)=\Phi_{\parallel}(t=0)=1$, significantly different from $\Phi_{\perp}(t=0)\neq\Phi_{\parallel}(t=0)$ here. Together with the analysis for 2D tilted Dirac bands \cite{arXivHou2022},
	this robust behavior of fixed point at $\omega=2\mu$ is universal for both 2D tilted SDBs and 2D tilted Dirac bands,
	in spite of different geometric structure of Fermi surface. Fourth, from the analytical expressions of $S_{\chi}^{(\mathrm{IB})}(\omega)$ and  $\Gamma_{\chi}^{(\mathrm{IB})}(\omega,\mu,t)$, two kinds of products $\mathrm{Re}\sigma_{\perp}^{\mathrm{IB}}(\omega=2\mu,\mu,0 < t\leq 2)
	\times\mathrm{Re}\sigma_{\parallel}^{\mathrm{IB}}(\omega=2\mu,\mu,0 < t\leq 2)$ and $\mathrm{Re}\sigma_{\perp}^{\mathrm{asymp}}(\omega,\mu,t)
	\times\mathrm{Re}\sigma_{\parallel}^{\mathrm{asymp}}(\omega,\mu,t)$ in 2D tilted SDBs are both different up to a factor $32/15\pi$ from their counterparts in 2D tilted Dirac bands \cite{PRBTan2022,arXivHou2022}. Fifth, the product $\mathrm{Re}\sigma_{\perp}^{\mathrm{asymp}}(\omega,\mu,t)\times \mathrm{Re}\sigma_{\parallel}^{\mathrm{asymp}}(\omega,\mu,t)$ is free of $\omega$ since the power-law scaling of $\omega$ cancels in $S_{\perp}^{(\mathrm{IB})}(\omega)\times S_{\parallel}^{(\mathrm{IB})}(\omega)$. In addition, the product $\mathrm{Re}\sigma_{\perp}^{\mathrm{asymp}}(\omega,\mu,t)\times \mathrm{Re}\sigma_{\parallel}^{\mathrm{asymp}}(\omega,\mu,t)$ is independent of tilt parameter $t$ when $0\leq t\leq 1$, while depends on band tilting in the type-II phase ($t>1$). Sixth, the angular dependence in 2D tilted SDBs differs significantly from that in 2D tilted Dirac bands \cite{arXivHou2022}, due mainly to the power-law scaling of $\omega$ in $S_{\chi}^{(\mathrm{IB})}(\omega)$.

It is not difficult to extend present study to the case when there is a gap parameter $\Delta$ in the tilted semi-Dirac model from the previous study in the untilted model as in \cite{PRBCarbotte2019,PRBCarbotte2019B}. For one thing, the gap parameter leads to a new kinked point in the undoped case, while makes no qualitative changes in the finite doped case \cite{PRBCarbotte2019}. For another, when the gap parameter is present with two tilted semi-Dirac points emerging on one valley, a new kinked point in the interband LOCs can be gained by the interband optical transition that relate the valence and conduction bands between two nodes \cite{PRBCarbotte2019B}. Behaviors of these new kinked points depending on tilt deserves a further study in another work. Last but not least, the band tilting along $k_x$ is also predicted. It turns out that the interband LOCs in the materials with band tilting along $k_x$, are acquired by substituting 
$\varepsilon_\kappa^\lambda(k_x,k_y)=\kappa t \sqrt{a} k_x+\lambda \sqrt{(a k_x^2)^2+(\hbar v_F k_y)^2}$ back into Eq. (\ref{interLOC}). Consequently, there is an amplification in magnitude of $\Gamma_{\chi}^{(\mathrm{IB})}(\omega,\mu,t)$ and a shift in the value of $\omega_{\pm}(t)$ due to the deformation of Fermi surface. In addition, the fixed point at $\omega=2\mu$ always exists for $0<t\le 2$, which further support that the fixed point at $\omega=2\mu$ is universal regardless of different geometric structure of Fermi surface. The power-law scaling of $\omega$ in $S_{\chi}^{(\mathrm{IB})}(\omega)$ still holds, i.e. $S_{\perp}^{(\mathrm{IB})}(\omega)\propto\sqrt{\omega}$ and $S_{\parallel}^{(\mathrm{IB})}(\omega)\propto1/\sqrt{\omega}$.

In summary, we theoretically investigated highly anisotropic optical conductivities in the type-I, type-II, and type-III phases of 2D tilted SDBs within linear response theory. This work presented characteristic optical signatures of 2D tilted SDBs. Our theoretical predictions are expected to be qualitatively valid for a large number of 2D tilted semi-Dirac materials and can be used to fingerprint 2D tilted SDBs from 2D untilted SDBs and 2D tilted Dirac bands in optical measurements.

	\section*{ACKNOWLEDGEMENTS\label{Sec:acknowledgements}}
	
	
	This work is partially supported by the National Natural Science Foundation of China under Grant Nos. 11547200 and 11874273. H.G. acknowledges financial support from NSERC of Canada and the FQRNT of the Province of Quebec.
	
	\appendix
	\allowdisplaybreaks[4]
	\begin{widetext}
		
		\section{Explicit definition of LOCs \label{App1}}

		We begin with the Hamiltonian in the vicinity of one of two valleys for 2D tilted semi-Dirac materials
		\begin{align}
			\mathcal{H}_\kappa(k_x,k_y)=\kappa v_t k_y\tau_0+a k_x^2\tau_1+v_Fk_y\tau_2,
		\end{align}
		where $\kappa=\pm$ labels two valleys, $\boldsymbol{k}=(k_x,k_y)$ stands for the wave vector,
		$\tau_0$ and $\tau_i$ denote the $2\times2$ unit matrix and Pauli matrices, respectively.
		The eigenvalue evaluated from the Hamiltonian reads
		\begin{align}
			\varepsilon_\kappa^\lambda(k_x,k_y)=\kappa v_t k_y+\lambda \mathcal{Z}(k_x,k_y),
		\end{align}
		where
		\begin{align}
			\mathcal{Z}(k_x,k_y)=\sqrt{\left(ak_x^2\right)^2+v_F^2 k_y^2},
		\end{align}
		and $\lambda=\pm1$ denotes the conduction and valence bands, respectively.
		
		Within the linear response theory, the longitudinal optical conductivities (LOCs) for the photon frequency $\omega$ and chemical potential $\mu$ is given by
		
		\begin{align}
			\sigma_{jj}(\omega,\mu,t)&=\sum_{\kappa=\pm}\sigma_{jj}^\kappa(\omega,\mu,t),
		\end{align}
		where the LOCs for chemical potential $\mu$ at the $\kappa$ valley can be expressed as
		\begin{align}
			\sigma_{jj}^{\kappa}(\omega,\mu,t)&=\frac{i}{\omega}
			\frac{1}{\beta}\sum_{\Omega_m}\int_{-\infty}^{+\infty}\frac{dk_x}{2\pi}\int_{-\infty}^{+\infty}\frac{dk_y}{2\pi}
			\mathrm{Tr}\left[\hat{J}_{j}^{\kappa}G_{\kappa}(k_x,k_y,i\Omega_m,\mu)
			\hat{J}_{j}^{\kappa}G_{\kappa}(k_x,k_y,i\Omega_m+\omega+i\eta,\mu)\right],
		\end{align}
		where $\beta=1/k_B T$, $j=x,y$ refer to spatial coordinates, $\eta$ denotes a positive infinitesimal. The charge
		current operators read
		\begin{align}
			\hat{J}_x^{\kappa}&=e\frac{\partial \mathcal{H}_\kappa(k_x,k_y)}{\partial k_x}=2ea k_x\tau_1,\\
			\hat{J}_y^{\kappa}&=e\frac{\partial \mathcal{H}_\kappa(k_x,k_y)}{\partial k_y}=e \left(\kappa v_t \tau_0+v_F\tau_2\right),
		\end{align}
		and the Matsubara Green's function in momentum space takes the form
		\begin{align}
			G_\kappa(k_x,k_y,i\Omega_m,\mu)
			&=\frac{1}{2}\sum_{\lambda=\pm}
			\frac{\mathcal{P}_{\kappa}^{\lambda}(k_x,k_y)}{i\Omega_m+\mu-\varepsilon_\kappa^\lambda(k_x,k_y)},
		\end{align}
		where $\mu$ is the chemical potential, and
		\begin{align}
			\mathcal{P}_{\kappa}^{\lambda}(k_x,k_y)
			&=\tau_0+\lambda\frac{ak_x^2 \tau_1+v_F k_y\tau_2}{\mathcal{Z}(k_x,k_y)}.
		\end{align}
		
		After summing over Matsubara frequency $\Omega_m$, we express the longitudinal optical conductivity for chemical potential $\mu$ at the $\kappa$ valley as
		\begin{align}
			\sigma_{jj}^{\kappa}(\omega,\mu,t)&
			=\frac{i}{\omega}\int_{-\infty}^{+\infty}\frac{dk_x}{2\pi}\int_{-\infty}^{+\infty}\frac{dk_y}{2\pi}
			\sum_{\lambda,\lambda^{\prime}=\pm}
			\mathcal{F}_{jj}^{\kappa;\lambda,\lambda^{\prime}}(k_x,k_y)
			\mathcal{M}_{\lambda,\lambda^{\prime}}^{\kappa}(k_x,k_y,\omega,\mu)
			\nonumber\\&
			= \frac{i}{\omega} \int_{-\infty}^{+\infty}\frac{dk_x}{2\pi} \int_{-\infty}^{+\infty}\frac{dk_y}{2\pi}\sum_{\lambda=\pm}\sum_{\lambda^\prime=\pm} \mathcal{F}_{jj}^{\kappa;\lambda,\lambda^\prime}(k_x,k_y)\frac{f\left[\varepsilon_\kappa^{\lambda}(k_x,k_y),\mu\right] -f\left[\varepsilon_\kappa^{\lambda^\prime}(k_x,k_y),\mu\right]}{\omega+\varepsilon_\kappa^{\lambda}(k_x,k_y)- \varepsilon_\kappa^{\lambda^\prime}(k_x,k_y)+i\eta},
			\label{DefOCApp}
		\end{align}
		where
		\begin{align}
			\mathcal{F}_{jj}^{\kappa;\lambda,\lambda^{\prime}}(k_x,k_y)
			=\frac{\mathrm{Tr}[\hat{J}_j^{\kappa}\mathcal{P}_{\kappa}^{\lambda}(k_x,k_y)
				\hat{J}_j^{\kappa}\mathcal{P}_{\kappa}^{\lambda^{\prime}}(k_x,k_y)]}{4},
		\end{align}
		and
		\begin{align}
			\mathcal{M}_{\lambda,\lambda^{\prime}}^{\kappa}(k_x,k_y,\omega,\mu)
			&=\frac{f[\varepsilon_{\kappa}^{\lambda}(k_x,k_y),\mu]
				-f[\varepsilon_{\kappa}^{\lambda^{\prime}}(k_x,k_y),\mu]} {\omega+\varepsilon_{\kappa}^{\lambda}(k_x,k_y)
				-\varepsilon_{\kappa}^{\lambda^{\prime}}(k_x,k_y)+i\eta},\label{MDef}
		\end{align}
		with $f(x,\mu)=1/\left\{1+\exp[\beta(x-\mu)]\right\}$ denoting the Fermi distribution function.
		
		Specifically, the explicit expressions of $\mathcal{F}_{jj}^{\kappa;\lambda,\lambda^{\prime}}(k_x,k_y)$
		are given as
		\begin{align}
			\mathcal{F}_{xx}^{\kappa;\lambda,\lambda^{\prime}}(k_x,k_y)
			&=\frac{\mathrm{Tr}[\hat{J}_x^{\kappa}\mathcal{P}_{\kappa}^{\lambda}(k_x,k_y)
				\hat{J}_x^{\kappa}\mathcal{P}_{\kappa}^{\lambda^{\prime}}(k_x,k_y)]}{4}
			\nonumber\\&
			=2e^2a^2k_x^2\left\{1+\lambda\lambda^{\prime}\frac{\left(ak_x^2\right)^2
				-v_F^2k_y^2}{\left[\mathcal{Z}(k_x,k_y)\right]^2}\right\},\label{FxxDef}
		\end{align}
		and
		\begin{align}
			\mathcal{F}_{yy}^{\kappa;\lambda,\lambda^{\prime}}(k_x,k_y)
			&=\frac{\mathrm{Tr}[\hat{J}_y^{\kappa}\mathcal{P}_{\kappa}^{\lambda}(k_x,k_y)
				\hat{J}_y^{\kappa}\mathcal{P}_{\kappa}^{\lambda^{\prime}}(k_x,k_y)]}{4}
			\nonumber\\&
			=\frac{e^2}{2}\left\{v_t^2(1+\lambda\lambda^{\prime})+v_F^2\left[1-\lambda\lambda^{\prime}\frac{\left(ak_x^2\right)^2
				-v_F^2k_y^2}{\left[\mathcal{Z}(k_x,k_y)\right]^2}\right]
			+2\kappa(\lambda+\lambda^{\prime})v_F^2\frac{v_tk_y}{\mathcal{Z}(k_x,k_y)}\right\}
			\nonumber\\&
			=\frac{e^2}{2}\left\{2v_t^2\delta_{\lambda\lambda^{\prime}}+v_F^2\left[1-\lambda\lambda^{\prime}\frac{\left(ak_x^2\right)^2
				-v_F^2k_y^2}{\left[\mathcal{Z}(k_x,k_y)\right]^2}\right]
			+4\kappa\lambda\delta_{\lambda\lambda^{\prime}}v_F^2\frac{v_tk_y}{\mathcal{Z}(k_x,k_y)}\right\}.\label{FyyDef}
		\end{align}

		It is easy to verify that $\sigma_{jj}(\omega,\mu,t)$ respects the particle-hole symmetry, namely,
		\begin{align}
			\sigma_{jj}(\omega,\mu,t)&=\sigma_{jj}(\omega,-\mu,t)=\sigma_{jj}(\omega,|\mu|,t).
		\end{align}
		
		Keeping this property of $\sigma_{jj}(\omega,\mu)$ in mind, we can safely replace $\mu$ in all of $\sigma_{jj}(\omega,\mu)$, $\sigma_{jj}^{\kappa}(\omega,\mu)$, and $f(x,\mu)$ by $\mu$ since we only concern the final result of $\sigma_{jj}(\omega,\mu)$. Hereafter, we restrict our analysis to the n-doped case ($\mu>0$).

		After some standard algebra, the real part of the LOCs can be divided into interband part and intraband part as
		\begin{align}
			\mathrm{Re}\sigma_{jj}^{\kappa}(\omega,\mu,t)
			=\begin{cases}
				\mathrm{Re}\sigma_{jj(\mathrm{IB})}^{\kappa}(\omega,\mu,t)+\Theta[\mu]\mathrm{Re}\sigma_{jj(\mathrm{D})}^{\kappa,+}(\omega,\mu,t)
				+\Theta[-\mu]\mathrm{Re}\sigma_{jj(\mathrm{D})}^{\kappa,-}(\omega,\mu,t), & 0\leq t \leq1,\\\\
				\mathrm{Re}\sigma_{jj(\mathrm{IB})}^{\kappa}(\omega,\mu,t)+\mathrm{Re}\sigma_{jj(\mathrm{D})}^{\kappa,+}(\omega,\mu,t)
				+\mathrm{Re}\sigma_{jj(\mathrm{D})}^{\kappa,-}(\omega,\mu,t), & t>1,
			\end{cases}
			\label{Eq4}
		\end{align}
		where $\Theta(x)$ is the Heaviside step function satisfying $\Theta(x)=0$ for $x\le0$ and $\Theta(x)=1$ for $x>0$, $\mu$ denotes the chemical potential measured with respect to the Dirac point, and the interband and intraband conductivities are given respectively as
		\begin{align}
			\mathrm{Re}\sigma_{jj(\mathrm{IB})}^\kappa(\omega,\mu,t)
			&=\pi \int^{+\infty}_{-\infty}\frac{dk_x}{2\pi} \int^{+\infty}_{-\infty}\frac{dk_y}{2\pi}\mathcal{F}^{\kappa;jj}_{-,+}(k_x,k_y) \frac{f\left[\varepsilon_\kappa^{-}(k_x,k_y),\mu\right] -f\left[\varepsilon_\kappa^{+}(k_x,k_y),\mu\right]} {\omega} \delta\left[\omega-2\mathcal{Z}(k_x,k_y)\right],\label{Eq5}\\
			\mathrm{Re}\sigma_{jj(\mathrm{D})}^{\kappa,\lambda}(\omega,\mu,t)
			&=\pi\int^{+\infty}_{-\infty}\frac{dk_x}{2\pi} \int^{+\infty}_{-\infty}\frac{dk_y}{2\pi} \mathcal{F}^{\kappa;jj}_{\lambda,\lambda}(k_x,k_y) \left[-\frac{d f\left[\varepsilon_\kappa^{\lambda}(k_x,k_y),\mu\right]} {d\varepsilon_\kappa^{\lambda}(k_x,k_y)}\right]\delta(\omega),
			\label{Eq6}
		\end{align}
		with $\delta(x)$ the Dirac $\delta$-function.

		\section{Detailed calculation of interband LOCs \label{App2}}
		
		In the next two sections, we will analytically calculate the interband and intraband LOCs by assuming zero temperature $T=0$ such that the Fermi distribution function $f(x)$ can be replaced by the Heaviside step function $\Theta[\mu-x]$. Consequently, we have
		\begin{align}
			\mathrm{Re}\sigma_{\perp(\mathrm{IB})}^\kappa(\omega,\mu,t)\equiv\mathrm{Re}\sigma_{xx(\mathrm{IB})}^\kappa(\omega,\mu,t)
			&=\pi \int^{+\infty}_{-\infty}\frac{dk_x}{2\pi} \int^{+\infty}_{-\infty}\frac{dk_y}{2\pi} \frac{1}{\omega}
			\left\{2e^2a^2k_x^2\left\{1-\frac{\left(ak_x^2\right)^2-v_F^2k_y^2}{\left(ak_x^2\right)^2+v_F^2k_y^2}\right\}\right\}
			\nonumber\\&\hspace{0.5cm}
			\left\{\Theta\left[\mu-\varepsilon_\kappa^{-}(k_x,k_y)\right]-\Theta\left[\mu-\varepsilon_\kappa^{+}(k_x,k_y)\right]\right\}
			\delta\left[\omega-2\sqrt{\left(ak_x^2\right)^2+v_F^2k_y^2}\right]\nonumber\\
			&=\pi\frac{2e^2a}{\omega}\lim_{\varepsilon \to 0^{+}}
			\left(-\frac{\partial}{\partial \varepsilon}\right)\left[\mathcal{T}_{\kappa}(\omega,\mu,\varepsilon)-\mathcal{L}_{\kappa}(\omega,\mu,\varepsilon)\right],\label{Eq9}
		\end{align}
		and
		\begin{align} \mathrm{Re}\sigma_{\parallel(\mathrm{IB})}^\kappa(\omega,\mu,t)\equiv\mathrm{Re}\sigma_{yy(\mathrm{IB})}^\kappa(\omega,\mu,t)
			&=\pi \int^{+\infty}_{-\infty}\frac{dk_x}{2\pi} \int^{+\infty}_{-\infty}\frac{dk_y}{2\pi} \frac{1}{\omega}
			\frac{e^2v_F^2}{2}\left[1+\frac{\left(ak_x^2\right)^2-v_F^2k_y^2}{\left(ak_x^2\right)^2+v_F^2k_y^2}\right]
			\nonumber\\&\hspace{0.5cm}
			\left\{\Theta\left[\mu-\varepsilon_\kappa^{-}(k_x,k_y)\right]-\Theta\left[\mu-\varepsilon_\kappa^{+}(k_x,k_y)\right]\right\}
			\delta\left[\omega-2\sqrt{\left(ak_x^2\right)^2+v_F^2k_y^2}\right]
			\nonumber\\&
			=\pi\frac{e^2v_F^2}{2\omega}\left[\mathcal{T}_{\kappa}(\omega,\mu,t,0)+\mathcal{L}_{\kappa}(\omega,\mu,t,0)\right],
			\label{Eq9}
		\end{align}
		where
		\begin{align}
			\mathcal{T}_{\kappa}(\omega,\mu,t,\varepsilon)
			=&\int_{-\infty}^{+\infty}\frac{dk_x}{2\pi}\int_{-\infty}^{+\infty}\frac{dk_y}{2\pi}
			\left\{\Theta\left[\mu-\varepsilon_\kappa^-(k_x,k_y)\right]- \Theta\left[\mu-\varepsilon_\kappa^+(k_x,k_y)\right]\right\}
			\nonumber\\&
			\delta\left[\omega-2\sqrt{\left(ak_x^2\right)^2+v_F^2k_y^2}\right]~\exp(-\varepsilon a k_{x}^{2}),
			\nonumber\\
			\mathcal{L}_{\kappa}(\omega,\mu,t,\varepsilon)
			=&\int_{-\infty}^{+\infty}\frac{dk_x}{2\pi}\int_{-\infty}^{+\infty}\frac{dk_y}{2\pi}
			\frac{\left(ak_x^2\right)^2-v_F^2k_y^2}{\left(ak_x^2\right)^2+v_F^2k_y^2}
			\left\{\Theta\left[\mu-\varepsilon_\kappa^-(k_x,k_y)\right]- \Theta\left[\mu-\varepsilon_\kappa^+(k_x,k_y)\right]\right\}
			\nonumber\\&
			\delta\left[\omega-2\sqrt{\left(ak_x^2\right)^2+v_F^2k_y^2}\right]~\exp(-\varepsilon a k_{x}^{2}).
		\end{align}
		
		After some simple algebra,  $\mathcal{T}_\kappa(\omega,\mu,t,\varepsilon)$ and $\mathcal{L}_{\kappa}(\omega,\mu,t,\varepsilon)$ can be written as
		\begin{align}
			\mathcal{T}_{\kappa}(\omega,\mu,t,\varepsilon)-\mathcal{L}_{\kappa}(\omega,\mu,t,\varepsilon)
			&=\frac{1}{\pi}\frac{\omega}{2e^2a}\sum_{n=0}^{+\infty}\frac{(-)^{n}}{n!}\varepsilon^{n}S_{xx}^{(\mathrm{IB})}(\omega,n)\Gamma_{xx}^{(\mathrm{IB})}(\omega,\mu,t,n),\\
			\mathcal{T}_{\kappa}(\omega,\mu,t,\varepsilon)+\mathcal{L}_{\kappa}(\omega,\mu,t,\varepsilon)
			&=\frac{1}{\pi}\frac{2\omega}{e^2v_F^2}\sum_{n=0}^{+\infty}\frac{(-)^{n}}{n!}\varepsilon^{n}S_{yy}^{(\mathrm{IB})}(\omega,n)\Gamma_{yy}^{(\mathrm{IB})}(\omega,\mu,t,n),
		\end{align}
		where $\Gamma_{jj}^{(\mathrm{IB})}(\omega,\mu,t,n)$ reads
		\begin{align}
			\Gamma_{xx}^{(\mathrm{IB})}(\omega,\mu,t,n)&=\int dy\left[(1-y^2)^{\frac{2n-3}{4}}-(1-y^2)^{\frac{2n+1}{4}}\right]\left[\Theta\left(\frac{\omega+2\mu}{t\omega}-y\right)-\Theta\left(y-\frac{\omega-2\mu}{t\omega}\right)\right],\\
			\Gamma_{yy}^{(\mathrm{IB})}(\omega,\mu,t,n)&=\int dy(1-y^2)^{\frac{2n+1}{4}}\left[\Theta\left(\frac{\omega+2\mu}{t\omega}-y\right)-\Theta\left(y-\frac{\omega-2\mu}{t\omega}\right)\right].
		\end{align}
		Therein, $\Gamma_{jj}^{(\mathrm{IB})}(\omega,\mu,t,n)$ behaves as a deformed Heaviside-like function, while $S_{jj}^{(\mathrm{IB})}(\omega,n)$ modulates its magnitudes with respect to different $\omega$ by
		\begin{align}
			S_{xx}^{(\mathrm{IB})}(\omega,n)&=\frac{e^2\sqrt{a}}{4\pi\upsilon_F}\left(\frac{\omega}{2}\right)^{\frac{2n-1}{2}},\\ S_{yy}^{(\mathrm{IB})}(\omega,n)&=\frac{e^2v_F}{16\pi\sqrt{a}}\left(\frac{\omega}{2}\right)^\frac{2n-1}{2}.
		\end{align}
		In order to express  $\Gamma_{jj}^{\mathrm{IB}}(\omega,\mu,t,n)$ in a more compact form, we introduce several useful notations
		\begin{align}
			\omega_{\pm}&=\frac{2\mu}{|1\mp t|},\\ \xi_{\pm}&=\frac{\omega\pm 2\mu}{t\omega}\frac{\Theta(t)}{t},\\
			\mathscr{B}(z,p,q)&=\frac{1}{2}\mathrm{sgn}(z)\mathrm{B}_{z^2}(p,q),
		\end{align}
		where $\mathrm{B}_{z}(p,q)$ is in general known as incomplete Beta function with $\mathrm{B}_{z}(p,q)=\int_0^{z}dx x^p(1-x)^{q-1}$ for $ z\in[0,1]$.
		
		For the type-I phase ($0<t<1$)
		\begin{align}
			\Gamma_{xx}^{(\mathrm{IB})}(\omega,\mu,0<t<1,n)=
			\begin{cases}
				0, & 0<\omega<\omega_{-},\\\\
				\mathscr{B}(1,\frac{1}{2},\frac{2n+1}{4})-\mathscr{B}(1,\frac{1}{2},\frac{2n+5}{4})
				\\+\mathscr{B}(\xi_{-},\frac{1}{2},\frac{2n+1}{4})-\mathscr{B}(\xi_{-},\frac{1}{2},\frac{2n+5}{4}), & \omega_{-}<\omega<\omega_{+},\\\\
				2\left[\mathscr{B}(1,\frac{1}{2},\frac{2n+1}{4})-\mathscr{B}(1,\frac{1}{2},\frac{2n+5}{4})\right], & \omega>\omega_{+},
			\end{cases},
		\end{align}
		and
		\begin{align}
			\Gamma_{yy}^{(\mathrm{IB})}(\omega,\mu,0<t<1,n)=
			\begin{cases}
				0, & 0<\omega<\omega_{-},\\\\
				\mathscr{B}(1,\frac{1}{2},\frac{2n+5}{4})+\mathscr{B}(\xi_{-},\frac{1}{2},\frac{2n+5}{4}), & \omega_{-}<\omega<\omega_{+},\\\\
				2\mathscr{B}(1,\frac{1}{2},\frac{2n+5}{4}), & \omega>\omega_{+}.
			\end{cases}
		\end{align}
		
		For the type-II phase ($t>1$),
		\begin{align}
			\Gamma_{xx}^{(\mathrm{IB})}(\omega,\mu,t>1,n)
			=&
			\begin{cases}
				0, & 0<\omega<\omega_{-},\\\\
				\mathscr{B}(1,\frac{1}{2},\frac{2n+1}{4})-\mathscr{B}(1,\frac{1}{2},\frac{2n+5}{4})
				\\+\mathscr{B}(\xi_{-},\frac{1}{2},\frac{2n+1}{4})-\mathscr{B}(\xi_{-},\frac{1}{2},\frac{2n+5}{4}), & \omega_{-}<\omega<\omega_{+},\\\\	
				\mathscr{B}(\xi_{-},\frac{1}{2},\frac{2n+1}{4})-\mathscr{B}(\xi_{-},\frac{1}{2},\frac{2n+5}{4})
				\\+\mathscr{B}(\xi_{+},\frac{1}{2},\frac{2n+1}{4})-\mathscr{B}(\xi_{+},\frac{1}{2},\frac{2n+5}{4}), & \omega>\omega_{+},
			\end{cases}
		\end{align}
		and
		\begin{align}
			\Gamma_{yy}^{(\mathrm{IB})}(\omega,\mu,t>1,n)
			&=
			\begin{cases}
				0, & 0<\omega<\omega_{-},\\\\
				\mathscr{B}(1,\frac{1}{2},\frac{2n+5}{4})+\mathscr{B}(\xi_{-},\frac{1}{2},\frac{2n+5}{4}), & \omega_{-}<\omega<\omega_{+},\\\\
				\mathscr{B}(\xi_{-},\frac{1}{2},\frac{2n+5}{4})+\mathscr{B}(\xi_{+},\frac{1}{2},\frac{2n+5}{4}), & \omega>\omega_{+}.
			\end{cases}
		\end{align}
		
		For the type-III phase ($t=1$),
		\begin{align}
			\Gamma_{xx}^{(\mathrm{IB})}(\omega,\mu,t=1,n)
			&=
			\begin{cases}
				0, & 0<\omega<\mu,\\\\
				\mathscr{B}(1,\frac{1}{2},\frac{2n+1}{4})-\mathscr{B}(1,\frac{1}{2},\frac{2n+5}{4})
				\\+\mathscr{B}(\xi_{-},\frac{1}{2},\frac{2n+1}{4})-\mathscr{B}(\xi_{-},\frac{1}{2},\frac{2n+5}{4}), & \omega>\mu,
			\end{cases}
		\end{align}
		and
		\begin{align}
			\Gamma_{yy}^{(\mathrm{IB})}(\omega,\mu,t=1,n)
			&=\begin{cases}
				0, & 0<\omega<\mu,\\\\
				\mathscr{B}(1,\frac{1}{2},\frac{2n+5}{4})+\mathscr{B}(\xi_{-},\frac{1}{2},\frac{2n+5}{4}), & \omega>\mu.
			\end{cases}
		\end{align}
		
		Hereafter, we should just input specific values for $n$ to acquire the interband LOCs in $\kappa$ as
		\begin{align}
			\mathrm{Re}\sigma^{\kappa}_{xx{(\mathrm{IB})}}(\omega,\mu,t)
			&=S_{xx}^{(\mathrm{IB})}(\omega,1)\Gamma_{xx}^{(\mathrm{IB})}(\omega,\mu,t,1),\\
			\mathrm{Re}\sigma^{\kappa}_{yy{(\mathrm{IB})}}(\omega,\mu,t)
			&=S_{yy}^{(\mathrm{IB})}(\omega,0)\Gamma_{yy}^{(\mathrm{IB})}(\omega,\mu,t,0).
		\end{align}
		
		Furthermore,
		\begin{align}
			&\mathrm{Re}\sigma_{xx}^{\mathrm{IB}}(\omega,\mu,t)=g_s\sum_{\kappa=\pm}\mathrm{Re}\sigma_{xx(\mathrm{IB})}^{\kappa}(\omega,\mu,t)=N_f S_{\perp}^{(\mathrm{IB})}(\omega)\Gamma_{\perp}^{(\mathrm{IB})}(\omega,\mu,t),\\
			&	 \mathrm{Re}\sigma_{yy}^{\mathrm{IB}}(\omega,\mu,t)=g_s\sum_{\kappa=\pm}\mathrm{Re}\sigma_{yy(\mathrm{IB})}^{\kappa}(\omega,\mu,t)=N_f S_{\parallel}^{(\mathrm{IB})}(\omega)\Gamma_{\parallel}^{(\mathrm{IB})}(\omega,\mu,t),
		\end{align}
		where $\sigma_{0}=\frac{e^2}{4\hbar}$ (we restore $\hbar$ for explicitness) and a degeneracy factor $N_f=g_s g_v$ accounting for spin degeneracy $g_s$ and valley degeneracy $g_v$. In the maintext,
		we denote $
		S_{\perp}^{(\mathrm{IB})}(\omega)=\frac{\sigma_{0}\sqrt{2a\omega}}{2\pi\upsilon_F}$ and $S_{\parallel}^{(\mathrm{IB})}(\omega)=\frac{\sigma_{0}v_F}{2\pi\sqrt{2a\omega}}$.
		
		\begin{table}[htbp]
			\begin{tabular*}
				{\columnwidth}{@{\extracolsep{\fill}} | c | c | c |}
				\hline
				&$\Gamma_{\perp}^{(\mathrm{IB})}(\omega,\mu,0<t<1)$&	 $\Gamma_{\parallel}^{(\mathrm{IB})}(\omega,\mu,0<t<1)$\\
				\hline
				$0<\omega<\omega_{-}$ & $0$ & $0$
				\\
				\hline
				$\omega_{-}\le\omega<\omega_{+}$& $\mathscr{B}(1,\frac{1}{2},\frac{3}{4})-\mathscr{B}(1,\frac{1}{2},\frac{7}{4})+\mathscr{B}(\xi_{-},\frac{1}{2},\frac{3}{4})-\mathscr{B}(\xi_{-},\frac{1}{2},\frac{7}{4})$ & $\mathscr{B}(1,\frac{1}{2},\frac{5}{4})+\mathscr{B}(\xi_{-},\frac{1}{2},\frac{5}{4})$
				\\
				\hline
				$\omega\ge\omega_{+}$& $2\left[\mathscr{B}(1,\frac{1}{2},\frac{3}{4})-\mathscr{B}(1,\frac{1}{2},\frac{7}{4})\right]$  &  $2\mathscr{B}(1,\frac{1}{2},\frac{5}{4})$
				\\
				\hline
			\end{tabular*}
			\caption{Explicit expressions of $\Gamma_{\perp}^{(\mathrm{IB})}(\omega,\mu,0<t<1)$ and $\Gamma_{\parallel}^{(\mathrm{IB})}(\omega,\mu,0<t<1)$.}
			\label{Tabs1}
		\end{table}
		
		\begin{table}[htbp]
			\begin{tabular*}
				{\columnwidth}{@{\extracolsep{\fill}} | c | c | c |}
				\hline
				&$\Gamma_{\perp}^{(\mathrm{IB})}(\omega,\mu,t>1)$&	 $\Gamma_{\parallel}^{(\mathrm{IB})}(\omega,\mu,t>1)$\\
				\hline
				$0<\omega<\omega_{-}$ & $0$ & $0$
				\\
				\hline
				$\omega_{-}\le\omega<\omega_{+}$& $\mathscr{B}(1,\frac{1}{2},\frac{3}{4})-\mathscr{B}(1,\frac{1}{2},\frac{7}{4})+\mathscr{B}(\xi_{-},\frac{1}{2},\frac{3}{4})-\mathscr{B}(\xi_{-},\frac{1}{2},\frac{7}{4})$ & $\mathscr{B}(1,\frac{1}{2},\frac{5}{4})+\mathscr{B}(\xi_{-},\frac{1}{2},\frac{5}{4})$
				\\
				\hline
				$\omega\ge\omega_{+}$& $\mathscr{B}(\xi_{-},\frac{1}{2},\frac{3}{4})-\mathscr{B}(\xi_{-},\frac{1}{2},\frac{7}{4})+\mathscr{B}(\xi_{+},\frac{1}{2},\frac{3}{4})-\mathscr{B}(\xi_{+},\frac{1}{2},\frac{7}{4})$  &  $	\mathscr{B}(\xi_{-},\frac{1}{2},\frac{5}{4})+\mathscr{B}(\xi_{+},\frac{1}{2},\frac{5}{4})$
				\\
				\hline
			\end{tabular*}
			\caption{Explicit expressions of $\Gamma_{\perp}^{(\mathrm{IB})}(\omega,\mu,t>1)$ and $\Gamma_{\parallel}^{(\mathrm{IB})}(\omega,\mu,t>1)$.}
			\label{Tabs2}
		\end{table}
		
		In the following, we list  $\Gamma_{\perp}^{(\mathrm{IB})}(\omega,\mu,t)$ and $\Gamma_{\parallel}^{(\mathrm{IB})}(\omega,\mu,t)$ for distinct values of $t$ in Tables \ref{Tabs1}, \ref{Tabs2}, and \ref{Tabs3}.
		
		\begin{table}[htbp]
			\begin{tabular*}
				{\columnwidth}{@{\extracolsep{\fill}} | c | c | c |}
				\hline
				&$\Gamma_{\perp}^{(\mathrm{IB})}(\omega,\mu,t=1)$&	 $\Gamma_{\parallel}^{(\mathrm{IB})}(\omega,\mu,t=1)$\\
				\hline
				$0<\omega<\mu$ & $0$ & $0$
				\\
				\hline
				$\omega\ge\mu$& $\mathscr{B}(1,\frac{1}{2},\frac{3}{4})-\mathscr{B}(1,\frac{1}{2},\frac{7}{4})
				+\mathscr{B}(\xi_{-},\frac{1}{2},\frac{3}{4})-\mathscr{B}(\xi_{-},\frac{1}{2},\frac{7}{4})$   &  $\mathscr{B}(1,\frac{1}{2},\frac{5}{4})+\mathscr{B}(\xi_{-},\frac{1}{2},\frac{5}{4})$
				\\
				\hline
			\end{tabular*}
			\caption{Explicit expressions of $\Gamma_{\perp}^{(\mathrm{IB})}(\omega,\mu,t=1)$ and $\Gamma_{\parallel}^{(\mathrm{IB})}(\omega,\mu,t=1)$.}
			\label{Tabs3}
		\end{table}

		\section{Detailed calculation of intraband LOCs \label{App4}}
		The real part of the intraband LOCs (or Drude conductivities) reads,
		\begin{align}
			\mathrm{Re}\sigma_{jj(D)}(\omega,\mu,t)
			=\begin{cases}
				g_s\sum\limits_{\kappa=\pm1}\Theta[\mu]\mathrm{Re}\sigma_{jj(\mathrm{D})}^{\kappa,+}(\omega,\mu,t)
				+g_s\sum\limits_{\kappa=\pm1}\Theta[-\mu]\mathrm{Re}\sigma_{jj(\mathrm{D})}^{\kappa,-}(\omega,\mu,t), & 0\leq t \leq1,\\\\
				g_s\sum\limits_{\kappa=\pm1}\mathrm{Re}\sigma_{jj(\mathrm{D})}^{\kappa,+}(\omega,\mu,t)
				+g_s\sum\limits_{\kappa=\pm1}\mathrm{Re}\sigma_{jj(\mathrm{D})}^{\kappa,-}(\omega,\mu,t), & t>1.
			\end{cases}
		\end{align}
		
		For arbitrary type of tilt, we can always perform the calculation of the terms like
		\begin{align}
			\mathrm{Re}\sigma_{jj(\mathrm{D})}^{\kappa,\lambda}(\omega,\mu,t)=\pi\int^{+\infty}_{-\infty}\frac{dk_x}{2\pi} \int^{+\infty}_{-\infty}\frac{dk_y}{2\pi} \mathcal{F}^{\kappa;jj}_{\lambda,\lambda}(k_x,k_y) \delta\left[\mu-\varepsilon^{\lambda}_{\kappa}(k_x,k_y)\right]\delta(\omega),
		\end{align}
		where we have replaced the derivative of Fermi distribution with  $\delta\left[\mu-\varepsilon^{\lambda}_{\kappa}(k_x,k_y)\right]$ at zero temperature.
		
		Explicitly, we have
		\begin{align}
			\mathrm{Re}\sigma_{xx(\mathrm{D})}^{\kappa,\lambda}(\omega,\mu,t)
			&=\pi\int^{+\infty}_{-\infty}\frac{dk_x}{2\pi} \int^{+\infty}_{-\infty}\frac{dk_y}{2\pi}2e^2a^2k_x^2\left[1+\frac{\left(ak_x^2\right)^2-v_F^2k_y^2}{\left(ak_x^2\right)^2+v_F^2k_y^2}\right]
			\delta\left(\omega\right)
			\sum_{\kappa=\pm}\delta\left[\mu-\varepsilon_\kappa^{\lambda}(k_x,k_y)\right]\nonumber\\
			&=\frac{4\pi e^2a}{4\pi^2\sqrt{a}v_F}\int^{+\infty}_{0}\frac{dx}{\sqrt{x}}\int^{+\infty}_{-\infty}d\tilde{k}_y  \frac{x^3}{x^2+\tilde{k}_y^2}\delta\left(\omega\right)\delta\left[\mu-t \tilde{k}_y-\lambda\sqrt{x^2+\tilde{k}_y^2}\right],
			\label{intraxx}
		\end{align}
		where $x=a k_x^2$, $\tilde{k}_y=v_F k_y$ and we have utilized $\mathrm{Re}\sigma_{jj(\mathrm{D})}^{\kappa,\lambda}(\omega,\mu,t)=\mathrm{Re}\sigma_{jj(\mathrm{D})}^{-\kappa,\lambda}(\omega,\mu)$ in the second line and given discussion restricted in $\kappa=+$. Similarly, $\mathrm{Re}\sigma_{yy(\mathrm{D})}^{\kappa,\lambda}(\omega,\mu)$ reads
		\begin{align}
			\mathrm{Re}\sigma_{yy(\mathrm{D})}^{\kappa,\lambda}(\omega,\mu,t)
			&=\pi\int^{+\infty}_{-\infty}\frac{dk_x}{2\pi} \int^{+\infty}_{-\infty}\frac{dk_y}{2\pi}\frac{e^2}{2}\left[2v_t^2+\frac{4\kappa\lambda v_F^2v_tk_y}{\sqrt{\left(ak_x^2\right)^2+v_F^2k_y^2}}+\frac{2v_F^4k_y^2}{\left(ak_x^2\right)^2+v_F^2k_y^2}\right]
			\delta\left(\omega\right)
			\delta\left[\mu-\varepsilon_\kappa^{\lambda}(k_x,k_y)\right]\nonumber\\
			&=\frac{2\pi e^2 v_F^2}{8\pi^2\sqrt{a}v_F}\int^{+\infty}_{0}\frac{dx}{\sqrt{x}}\int^{+\infty}_{-\infty}d\tilde{k}_y  \left[t^2+\frac{2\lambda t\tilde{k}_y}{\sqrt{x^2+\tilde{k}_y^2}}+\frac{\tilde{k}_y^2}{x^2+\tilde{k}_y^2}\right]\delta\left(\omega\right)\delta\left[\mu-t \tilde{k}_y-\lambda\sqrt{x^2+\tilde{k}_y^2}\right].
			\label{intrayy}
		\end{align}
		
		Hence, we could focus only on Eqs.(\ref{intraxx}) and (\ref{intrayy}) to calculate intraband LOCs for arbitrary tilt. In order to simplify the results, the first Appell function
		\begin{align}
			F_1\left(a;b_1,b_2;c;x,y\right)=\frac{1}{\mathrm{B}\left(a,c-a\right)}\int^{1}_{0}t^{a-1}(1-t)^{c-a-1}(1-xt)^{-b_1}(1-yt)^{-b_2}dt, ~~~~~\mathrm{Re}(c)>\mathrm{Re}(a)>0,
		\end{align}
		is introduced in our result with its integral definition.
		To present our results in a more elegant fashion, we introduce some auxiliary functions with the help of  $\tilde{\omega}_{\pm}=\omega_{\pm}(t)/\mu$, $\tilde{\Lambda}=\Lambda/\mu$, and
		\begin{align}
			\begin{cases}
				\mathrm{S}_{\perp}^{(\mathrm{D})}(\mu)=\frac{\sigma_{0}\mu}{2\pi}\frac{\sqrt{a\mu}}{\upsilon_F},~~~\mathrm{S}_{\parallel}^{(\mathrm{D})}(\mu)=\frac{\sigma_{0}\mu}{2\pi}\frac{\upsilon_F}{\sqrt{a\mu}},\\\\
				X_u^\lambda(0<t<1)=\frac{\tilde{\omega}_{-}+\tilde{\omega}_{+}}{1+t\lambda \tilde{\omega}_{\lambda}},~~~X_o^\lambda(t>1,\tilde{\Lambda})=\frac{\tilde{\Lambda}+\lambda\tilde{\omega}_{-\lambda}}{\lambda-t\lambda\tilde{\omega}_{-\lambda}},~~~Y_o^\lambda(t>1,\tilde{\Lambda})=\frac{\tilde{\Lambda}+\lambda\tilde{\omega}_{-\lambda}}{\tilde{\omega}_{+}-\tilde{\omega}_{-}},\\\\
				\mathscr{F}^{\mathrm{I}}\left(a;b_1,b_2;c;x,y\right)=\frac{x~y^{-\tilde{\Theta}(t-1)}}{\tilde{\omega}_{+}+\mathrm{sgn}(1-t)\tilde{\omega}_{-}}t^{-3}~ F_{1}\left(a;b_1,b_2;c;x,y\right),\\\\
				\mathscr{F}^{\mathrm{II}}\left(a;b_1,b_2;c;x,y\right)=\frac{\tilde{\omega}_{+}+\mathrm{sgn}(1-t)\tilde{\omega}_{-}}{x~y^{-\tilde{\Theta}(t-1)}}t^3\left(1-t^{-2}\right)^2~F_{1}\left(a;b_1,b_2;c;x,y\right),\\\\
				\mathscr{F}^{\mathrm{III}}\left(a;b_1,b_2;c;x,y\right)=2\left(1-t^{-2}\right)~F_{1}\left(a;b_1,b_2;c;x,y\right),\\\\
				\mathcal{X}\left(\lambda;a;b_1,b_2;c;x,y\right)=8\lambda t^2\left|1-t^2\right|^{3/4}\left(\tilde{\omega}_{+}+\mathrm{sgn}(1-t)\tilde{\omega}_{-}\right)^{5/2}\mathscr{F}^{\mathrm{I}}\left(a;b_1,b_2;c;x,y\right),\\\\			 \mathcal{Y}\left(\lambda;a;b_1,b_2;c;x,y\right)=\frac{2\left|1-t^2\right|^{-3/4}\left[
					\lambda\mathscr{F}^{\mathrm{I}}\left(a;b_1,b_2;c;x,y\right)+\lambda\mathscr{F}^{\mathrm{II}}\left(a;-b_1,b_2;c;x,y\right)+\mathscr{F}^{\mathrm{III}}\left(a,-b_1,b_2;c;0,y\right)\right]}{\left(\tilde{\omega}_{+}+\mathrm{sgn}(1-t)\tilde{\omega}_{-}\right)^{1/2}}.
			\end{cases}
		\end{align}
		
		\subsection{calculation of $\mathrm{Re}\sigma_{jj}^{\mathrm{D}}(\omega,\mu,t)$ in the type-I phase ($0<t<1$)}
		In this subsection, we present details of calculation for intraband LOCs within $0<t<1$, and some tricks displayed in following could also be utilized in the calculation within $t>1$. The following calculations are allowed to  restrict to $\mathrm{Re}\sigma_{jj(\mathrm{D})}^{\kappa,\lambda}$, as the LOCs for the $\kappa$ valley is identical to it for the $-\kappa$ valley. It follows that
		\begin{align}
			&\mathrm{Re}\sigma_{xx(\mathrm{D})}^{\kappa,\lambda}(\omega,\lambda\mu,t)\nonumber\\
			&=\frac{4\pi e^2a}{4\pi^2\sqrt{a}v_F}\int^{+\infty}_{0}\frac{dx}{\sqrt{x}}\int^{+\infty}_{-\infty}d\tilde{k}_y  \frac{x^3}{x^2+\tilde{k}_y^2}\delta\left(\omega\right)\delta\left[\lambda\mu-t \tilde{k}_y-\lambda\sqrt{x^2+\tilde{k}_y^2}\right]\nonumber,\\
			&=\frac{4\pi e^2a}{4\pi^2\sqrt{a}v_F}\int^{+\infty}_{-\infty}d\tilde{k}_y \int^{+\infty}_{0} dx \frac{x^{\frac{5}{2}}}{x^2+\tilde{k}_y^2}\frac{\delta(x-x_0)+\delta(x+x_0)}{\left|\frac{x_0}{\mu-t\lambda\tilde{k}_y}\right|}\Theta\left[\left(\mu-t\lambda\tilde{k}_y\right)^2-\tilde{k}_y^2\right]\Theta(\mu-t\lambda\tilde{k}_y)\delta\left(\omega\right)\nonumber\\
			&= \mathrm{S}_{\perp}^{(\mathrm{D})}(\mu)\delta(\omega)\mathrm{B}\left(\frac{7}{4},\frac{7}{4}\right)\mathcal{X}\left(\lambda;\frac{7}{4};1,1;\frac{7}{2};t\lambda X_u^{\lambda}(t),0\right),
		\end{align}
		where $x_0=\sqrt{\left(\mu-\tilde{k}_y\right)^2-\tilde{k}_y^2}$ and $y=\frac{\tilde{k}_y}{\mu}$.
		Similarly, the $yy$-component of intraband LOCs in the type-I phase ($0<t<1$),
		\begin{align}
			&\mathrm{Re}\sigma_{yy(\mathrm{D})}^{\kappa,+}(\omega,\lambda\mu,t)\nonumber\\
			&=\frac{2\pi e^2 v_F^2}{8\pi^2\sqrt{a}v_F}
			\int^{+\infty}_{0}\frac{dx}{\sqrt{x}}\int^{+\infty}_{-\infty}d\tilde{k}_y  \left[t^2+\frac{2t\lambda\tilde{k}_y}{\sqrt{x^2+\tilde{k}_y^2}}+\frac{\tilde{k}_y^2}{x^2+\tilde{k}_y^2}\right]\delta\left(\omega\right)\delta\left[\lambda\mu-t \tilde{k}_y-\lambda\sqrt{x^2+\tilde{k}_y^2}\right]\nonumber\\
			&=\frac{2\pi e^2 v_F^2}{8\pi^2\sqrt{a}v_F}
			\int^{+\infty}_{-\infty}d\tilde{k}_y \int^{+\infty}_{0} \frac{dx}{\sqrt{x}}\left[t^2+\frac{2t\lambda\tilde{k}_y}{\sqrt{x^2+\tilde{k}_y^2}}+\frac{\tilde{k}_y^2}{x^2+\tilde{k}_y^2}\right]  \frac{\delta(x-x_0)+\delta(x+x_0)}{\left|\frac{x_0}{\mu-t\lambda\tilde{k}_y}\right|} \nonumber\\
			&\hspace{8cm}\Theta\left[\left(\mu-t\lambda\tilde{k}_y\right)^2-\tilde{k}_y^2\right]\Theta(\mu-t\lambda\tilde{k}_y)\delta\left(\omega\right)\nonumber\\
			&=\mathrm{S}^{(\mathrm{D})}_{\parallel}(\mu)\delta(\omega)\mathrm{B}\left(\frac{1}{4},\frac{1}{4}\right)\mathcal{Y}\left(\lambda;\frac{1}{4} ;1,1;\frac{1}{2};t\lambda X_u^\lambda(t),0\right).
		\end{align}
		As a consequence, we arrive at
		\begin{align}
			\mathrm{Re}\sigma_{xx}^{\mathrm{D}}(\omega,\left|\mu\right|,t)=&g_s\sum\limits_{\kappa=\pm1}\mathrm{Re}\sigma_{xx(\mathrm{D})}^{\kappa,\lambda=+}(\omega,\left|\mu\right|)
			=N_f\mathrm{S}_{\perp}^{(\mathrm{D})}(\mu)\delta(\omega)\mathrm{B}\left(\frac{7}{4},\frac{7}{4}\right)\mathcal{X}\left(+;\frac{7}{4};1,1;\frac{7}{2};t X_u^{+}(t),0\right),
		\end{align}
		and
		\begin{align}
			\mathrm{Re}\sigma_{yy}^{\mathrm{D}}(\omega,\left|\mu\right|)=&\sum\limits_{\kappa=\pm1}\mathrm{Re}\sigma_{yy(\mathrm{D})}^{\kappa,\lambda=+}(\omega,\left|\mu\right|,t)
			=N_f\mathrm{S}^{(\mathrm{D})}_{\parallel}(\mu)\delta(\omega)\mathrm{B}\left(\frac{1}{4},\frac{1}{4}\right)\mathcal{Y}\left(+;\frac{1}{4} ;1,1;\frac{1}{2};t X_u^+(t),0\right).
		\end{align}
		It is noted that only conduction band ($\lambda=+$) contributes to intraband LOCs with $\mu>0$.
		
		\subsection{calculation of $\mathrm{Re}\sigma_{jj}^{\mathrm{D}}(\omega,\mu,t)$ in the type-II phase ($t>1$)}
		The calculation of intraband LOCs for $t>1$ can be performed in a parallel way as for $0<t<1$. In the following, there are two points to emphasize. First, there are two Fermi surface, and hence we sum over $\lambda=\pm$ to acquire contributions from conduction band ($\lambda=+$) and valence band ($\lambda=-$). Second, $\tilde{\Lambda}$ accounts for the cutoff of integral in large $y$, i.e., $0<\upsilon_F k_y/\mu<\tilde{\Lambda}$. As a consequence, we arrive at
		\begin{align}
			&\mathrm{Re}\sigma_{xx}^{\mathrm{D}}(\omega,\mu>0,t)=g_s\sum\limits_{\kappa=\pm1}\sum\limits_{\lambda=\pm1}\mathrm{Re}\sigma_{xx(\mathrm{D})}^{\kappa,\lambda}(\omega,\mu>0,t)\nonumber\\
			=&N_f\mathrm{S}^{(\mathrm{D})}_{\perp}(\mu)\delta(\omega)\sum\limits_{\lambda=\pm1}\lambda\mathrm{B}\left(\frac{7}{4},1\right)\left[Y_o^{\lambda}(t,\tilde{\Lambda})\right]^{7/4}\mathcal{Y}\left(\lambda;\frac{7}{4};1,-\frac{3}{4};\frac{11}{4};-tX_o^{\lambda}(t,\tilde{\Lambda}),-Y_o^{\lambda}(t,\tilde{\Lambda})\right),
		\end{align}
		and
		\begin{align}
			&\mathrm{Re}\sigma_{yy}^{\mathrm{D}}(\omega,\mu>0,t)=g_s\sum\limits_{\kappa=\pm1}\sum\limits_{\lambda=\pm1}\mathrm{Re}\sigma_{yy(\mathrm{D})}^{\kappa,\lambda}(\omega,\mu>0,t)\nonumber\\
			=&N_f S^{(\mathrm{D})}_{\parallel}(\mu)\delta(\omega)\sum\limits_{\lambda=\pm1}\lambda \mathrm{B}\left(\frac{1}{4},1\right)\left[Y_o^{\lambda}(t,\tilde{\Lambda})\right]^{1/4}\mathcal{Y}\left(\lambda;\frac{1}{4};1,\frac{3}{4};\frac{7}{4};-tX_o^{\lambda}(t,\tilde{\Lambda}),-Y_o^{\lambda}(t,\tilde{\Lambda})\right).
		\end{align}
		
		\subsection{calculation of $\mathrm{Re}\sigma_{jj}^{\mathrm{D}}(\omega,\mu,t)$ in the type-III phase ($t=1$)}
		Physically, states occupying $(\kappa=\pm,\lambda=-)$ with energy $\varepsilon_{\kappa}^-=(k_x,k_y)$ are far below the Fermi surface in type-III phase, which leads to a negligible contribution to the intraband LOCs. Hence, $g_s\sum\limits_{\kappa=\pm1}\mathrm{Re}\sigma_{jj(\mathrm{D})}^{\kappa,-}(\omega,\mu>0,t)$ would vanish at $t=1$. As a result, we are allowed to focus only on $g_s\sum\limits_{\kappa=\pm1}\mathrm{Re}\sigma_{jj(\mathrm{D})}^{\kappa,+}(\omega,\mu>0,t)$, which gives rise to
		\begin{align}
			\mathrm{Re}\sigma_{xx}^{\mathrm{D}}(\omega,\mu,t)
			=g_s\sum\limits_{\kappa=\pm1}\mathrm{Re}\sigma_{xx(\mathrm{D})}^{\kappa,+}(\omega,\mu,t)
			&=N_f S^{(\mathrm{D})}_{\parallel}(\mu)\delta\left(\omega\right)8\left(1+2\tilde{\Lambda}\right)^{3/4}\frac{1+2\tilde{\Lambda}}{2+2\tilde{\Lambda}}\mathrm{B}\left(1,1\right) F_{1}\left(1;-\frac{3}{4},1;2;1,\frac{1+2\tilde{\Lambda}}{2+2\tilde{\Lambda}}\right),
		\end{align}
		and
		\begin{align}
			\mathrm{Re}\sigma_{yy}^{\mathrm{D}}(\omega,\mu,t)=g_s\sum\limits_{\kappa=\pm1}\mathrm{Re}\sigma_{yy(\mathrm{D})}^{\kappa,+}(\omega,\mu,t)
			&=N_f S^{(\mathrm{D})}_{\perp}(\mu)\delta\left(\omega\right)2(1+2\tilde{\Lambda})^{-3/4}\frac{1+2\tilde{\Lambda}}{2+2\tilde{\Lambda}}\mathrm{B}\left(1,1\right) F_{1}\left(1;\frac{3}{4},1;2;1,\frac{1+2\tilde{\Lambda}}{2+2\tilde{\Lambda}}\right).
		\end{align}
	\end{widetext}


\end{document}